\documentclass[preprint]{aastex}
\usepackage{amsbsy}

\newcommand{\eqref}[1]{(\ref{#1})}
\renewcommand{\vec}[1]{\boldsymbol{#1}}
\newcommand{\tensor}[1]{\mathbf{#1}}

\shorttitle{Planetesimal Dynamics in a Turbulent Disk}
\shortauthors{Yang, Mac Low, \& Menou}

\begin{document}

\title{Planetesimal and Protoplanet Dynamics
       in a Turbulent Protoplanetary Disk:\\
       Ideal Unstratified Disks}

\author{Chao-Chin Yang}
\affil{Department of Astronomy, University of Illinois, Urbana, IL 61801}
\affil{Department of Astrophysics, American Museum of Natural History,
       New York, NY 10024}
\email{cyang@amnh.org}

\author{Mordecai-Mark Mac Low}
\affil{Department of Astrophysics, American Museum of Natural History,
       New York, NY 10024}
\email{mordecai@amnh.org}

\and

\author{Kristen Menou}
\affil{Department of Astronomy, Columbia University, New York, NY 10027}
\email{kristen@astro.columbia.edu}

\begin{abstract}
The dynamics of planetesimals and planetary cores may be strongly influenced by density perturbations driven by magneto-rotational turbulence in their natal protoplanetary gas disks.  Using the local shearing box approximation, we perform numerical simulations of planetesimals moving as massless particles in a turbulent, magnetized, unstratified gas disk.  Our fiducial disk model shows turbulent accretion characterized by a Shakura-Sunyaev viscosity parameter of $\alpha \sim 10^{-2}$, with root-mean-square density perturbations of $\sim$10\%.  We measure the statistical evolution of particle orbital properties in our simulations including mean radius, eccentricity, and velocity dispersion.  We confirm random walk growth in time of all three properties, the first time that this has been done with direct orbital integration in a local model.  We find that the growth rate increases with the box size used at least up to boxes of eight scale heights in horizontal size.  However, even our largest boxes show velocity dispersions sufficiently low that collisional destruction of planetesimals should be unimportant in the inner disk throughout its lifetime.  Our direct integrations agree with earlier torque measurements showing that type~I migration dominates over diffusive migration by stochastic torques for most objects in the planetary core and terrestrial planet mass range.  Diffusive migration remains important for objects in the mass range of kilometer-sized planetesimals.  Discrepancies in the derived magnitude of turbulence between local and global simulations of magneto-rotationally unstable disks remains an open issue, with important consequences for planet formation scenarios.
\end{abstract}

\keywords{Accretion, accretion disks
      --- methods: numerical
      --- MHD
      --- planetary systems: formation
      --- planetary systems: protoplanetary disks
      --- turbulence}

\section{INTRODUCTION} \label{S:intro}

The core accretion scenario is a key ingredient of our current theory for planet formation \citep[see reviews by][and references therein]{jL93,PT06}.  In this model, solid bodies ranging from micron-sized dust grains to kilometer-sized planetesimals are the building blocks from which planetary cores assemble and ultimately form giant planets orbiting stars.  The dynamics of planetesimals and planetary cores may be strongly influenced by their natal protoplanetary gas disks.  In particular, type~I migration due to angular momentum exchange with the gas disk could drive a planetary core into the host star in a short timescale \citep[e.g.,][]{GT80,wW97,TTW02,MG04} compared to a typical disk lifetime of $\lesssim$10~Myr \citep[e.g.,][]{HC98,aS06,lH08}.  This remains one of the major obstacles to planet formation in the core accretion scenario unless recent models of the countervailing effects of type~III migration in non-adiabatic disks are confirmed \citep{PM06,KC08,PP09}.

Turbulent transport in young protoplanetary disks seems required for mass accretion to occur at observed rates.  The most likely cause of this turbulence is that disks are unstable to linear perturbations if they are weakly magnetized \citep{BH91}.  Although this magneto-rotational instability (MRI) is the most promising mechanism to drive turbulence in protoplanetary disks, the numerical convergence of simulations with increasing resolution continues to be an open issue.  It has been shown that MRI-driven turbulence in ideal, magnetized, unstratified disks without mean magnetic fields decreases with increasing resolution and might be negligible when resolution is high and numerical dissipation is small \citep{FP07a,PCP07}.  To maintain a nonzero, convergent level of turbulent accretion, it has been argued that simulated disks might require stratification \citep{DSP09,SKH09}, explicit dissipation \citep[e.g.,][]{FP07b,LL07}, or nonzero magnetic flux \citep[e.g.,][]{GG09,SHB09}.  Furthermore, the conductivity near the mid-plane of a protoplanetary disk may be too low for the MRI to operate, at least over some range of radii, resulting in a quiescent region known as a dead zone \citep[e.g.,][]{cG96,SM00,FTB02,SWH04,IN06,TSD07}.  Disks containing dead zones will, nevertheless, be stirred by turbulent regions near the disk surface \citep{FS03,OM09}.

The turbulent nature of protoplanetary disks might lead to scenarios for the formation and migration of planetesimals and planetary cores that are rather different from those suggested for the better-studied case of a laminar disk.  The turbulence driven by the MRI causes density enhancements significant enough to exert gravitational torques that turn the orbital motion of low-mass objects into a random walk (\citealt[hereafter LSA04]{LSA04}; \citealt{NP04}).  It has been shown that radial excursions and eccentricities of planetesimals or protoplanets could be excited through such a process (\citealt[hereafter N05]{rN05}; \citealt[hereafter OIM07]{OIM07}).  Since some protoplanets could diffuse their way radially outward in the process, type~I migration might be effectively delayed and some objects might more easily survive past the gas disk depletion (\citealt[hereafter JGM06]{JGM06}; \citealt{AB09}).  Even if the disk contains a dead zone, \citet[hereafter OMM07]{OMM07} have shown that low-mass objects within the dead zone still experience some of the turbulent torques generated by the active layers.  \citet[hereafter IGM08]{IGM08} suggest, on the other hand, that the velocity dispersion of planetesimals excited by hydromagnetic turbulence might be so strong that kilometer-sized objects suffer from collisional destruction.

The survivability of planetesimals or protoplanets under type~I migration or collisional destruction sensitively depends on their orbital dynamics in a turbulent gas disk.  Previous direct orbital integrations of planetesimals or protoplanets embedded in a turbulent gas disk were conducted in global disk models (LSA04; N05; OIM07; IGM08).  In contrast to global disk models, it can be advantageous to employ the local shearing box approximation \citep[e.g.,][]{GL65,HGB95,BN95} because of its high resolving power on turbulence structures and the possibility of integrating for long times.  \citet{NP04} and OMM07 first measured the stochastic torques generated by hydromagnetic turbulence at the center of a local shearing box.  In this paper, we pursue this topic further by using direct orbital integration of planetesimals moving as massless particles under the gravitational influence of MRI-driven turbulence in a local shearing box.  We focus our attention on unstratified disks in the context of ideal magnetohydrodynamics (MHD).  To maintain a nonzero, numerically convergent level of stochastic perturbations driven by the MRI, we impose a constant vertical magnetic flux.  We describe our numerical models in \S\ref{S:nm} and present the simulations in \S\ref{S:sr}, along with statistical analyses of the disk properties and the planetesimal orbits.  In \S\ref{S:ipf}, we use our results to revisit the issue of survivability of planetesimals and planetary cores, before reaching our conclusions in \S\ref{S:conc}.

\section{NUMERICAL MODELING} \label{S:nm}

We use the parallelized, cache-efficient, Pencil Code described by \citet{BD02}.  It solves the non-ideal MHD equations by sixth-order finite differences in space and third-order Runge-Kutta steps in time.  The induction equation is solved using the magnetic vector potential $\vec{A}$ so that zero divergence of magnetic field $\vec{B}$ is guaranteed at all time.  To save memory usage, the Runge-Kutta time integration is performed using the $2N$-method \citep{jW80}.  The scheme is not written in conservative form.  Instead, conserved quantities like total mass are monitored to evaluate the quality of the solution.  In the following subsections, we describe the equations assumed in our models as well as the numerical constructs.

  \subsection{Magnetohydrodynamics} \label{SS:mhd}
  
We use the local shearing box approximation \citep[e.g.,][]{GL65,BN95,HGB95} to simulate a small Cartesian box carved out of a Keplerian disk at a large distance from the host star.  The center of the box co-rotates with the disk at Keplerian angular speed $\Omega_K$, the $x$-axis is directed radially, and the $y$-axis is directed azimuthally.  The vertical component of gravity from the host star is ignored and thus the disk is unstratified.  We impose a vertical, external magnetic field $\vec{B}_\mathrm{ext} = B_\mathrm{ext}\hat{\vec{z}}$ to maintain a finite magnetic flux.  The MHD equations then become
\begin{eqnarray} \label{E:mhd}
  \partial_t\rho
  - \frac{3}{2}\Omega_K x\partial_y\rho
  + \nabla\cdot(\rho\vec{u})
  &=& f_D
  + \nabla\cdot\left(\nu_s\nabla\rho\right),\label{E:mhd_cont}\\
  \partial_t\vec{u}
  - \frac{3}{2}\Omega_K x\partial_y\vec{u}
  + \vec{u}\cdot\nabla\vec{u}
  &=& -\frac{1}{\rho}\nabla p
  + \left(2\Omega_K u_y\hat{\vec{x}}
          - \frac{1}{2}\Omega_K u_x\hat{\vec{y}}\right)
  + \frac{1}{\rho}\vec{J}\times\left(\vec{B} + 
                                     \vec{B}_\mathrm{ext}\right)\nonumber\\
  &&+ \vec{f}_V
  + \frac{1}{\rho}\nabla\left(\nu_s\rho\nabla\cdot\vec{u}\right),
    \label{E:mhd_mom}\\
  \partial_t\vec{A}
  - \frac{3}{2}\Omega_K x\partial_y\vec{A}
  &=& \frac{3}{2}\Omega_K A_y\hat{\vec{x}}
  + \vec{u}\times\left(\vec{B} + \vec{B}_\mathrm{ext}\right)
  + \vec{f}_R - \eta_s\vec{J},\label{E:mhd_ind}
\end{eqnarray}
in which $\rho$ is gas density, $\vec{u}$ is gas velocity relative to the background shear flow, $p$ is gas pressure, $\vec{J} = \nabla\times\vec{B} / \mu_0$ is the electric current density, $\vec{B} = \nabla\times\vec{A}$, and $\mu_0$ is permeability.  The terms $f_D$, $\vec{f}_V$, and $\vec{f}_R$, and those containing scalar variables $\nu_s$ and $\eta_s$ are numerical dissipation terms needed to stabilize the scheme, which are described below.  They are needed to resolve shocks, and because the difference scheme formally has vanishing dissipation.  We assume an isothermal equation of state, $p = c_s^2\rho$, where $c_s$ is the isothermal speed of sound.

The mass diffusion term $\nabla\cdot\left(\nu_s\nabla\rho\right)$ in the continuity equation~\eqref{E:mhd_cont} and the bulk viscosity term $\nabla\left(\nu_s\rho\nabla\cdot\vec{u}\right) / \rho$ in the momentum equation~\eqref{E:mhd_mom} are implemented to broaden shocks.  The artificial kinematic viscosity $\nu_s$ is of von Neumann type \citep[c.f.,][]{HBM04,LJ08}:
\begin{equation}
  \nu_s = \left\{
  \begin{array}{ll}
    -h^2\nabla\cdot\vec{u}, & \mathrm{if}\quad\nabla\cdot\vec{u} <
                                 -c_s / 4h,\\
    0,                         & \mathrm{otherwise,}
  \end{array}\right.
\end{equation}
where $h$ is grid spacing.  It is smoothed by taking a maximum over nearest neighbors and then convolved with a Gaussian kernel having a standard deviation of $h$.  Note that the threshold for the velocity divergence is set for eliminating artificial diffusion where hydrodynamic shocks are unlikely to be present.

We also include the Ohmic term $-\eta_s\vec{J}$ in the induction equation~\eqref{E:mhd_ind} to broaden strong current sheets.  The artificial resistivity $\eta_s$ assumes the same form used by \citet{NT01} but with a lower cutoff:
\begin{equation}
  \eta_s = \left\{
  \begin{array}{ll}
    \mu_0 h v_A, & \mathrm{if}\quad v_A > 8c_s,\\
    0,           & \mathrm{otherwise,}
  \end{array}\right.\label{E:nr}
\end{equation}
where $v_A = \left|\vec{B} + \vec{B}_\mathrm{ext}\right| / \sqrt{\mu_0\rho}$ is the Alv\'{e}n speed.  Although simple, this form may perform better to resolve sharp magnetic structures than a resistivity proportional to the magnitude of current density itself \citep{FA05}.  We apply a lower cutoff in equation~\eqref{E:nr} to only treat regions where fast magnetic reconnection may occur.

In addition to applying artificial diffusions addressing shocks and current sheets, we also implement hyper-diffusion $f_D$, hyper-viscosity $\vec{f}_V$, and hyper-resistivity $\vec{f}_R$ in the respective MHD equations~\eqref{E:mhd_cont}--\eqref{E:mhd_ind} \citep[e.g.,][]{HB04,JK05}:
\begin{equation}
  f_D          = \nu_3\nabla^6\rho,\quad
  \vec{f}_V = \nu_3\left(\nabla^6\vec{u} + \tensor{S}\cdot\nabla\ln\rho\right),
  \quad\mathrm{and}\quad
  \vec{f}_R = \nu_3\nabla^6\vec{A},
\end{equation}
where $\nu_3$ is a constant, the tensor $\tensor{S}$ is defined by $S_{ij} \equiv \partial_j^5 u_i$, and the sixth-order differential operator $\nabla^6 \equiv \partial_x^6 + \partial_y^6 + \partial_z^6$.  These terms are included in order to stabilize the high-order finite-difference scheme implemented by the Pencil Code.  The corresponding diffusivity is proportional to $k^6$, where $k$ is the wavenumber of a signal in the simulation, so features at small scales dissipate at a much higher rate than those at larger scales.  Therefore, by adjusting the coefficient $\nu_3$, we can maintain numerical stability by damping oscillations near the grid scale while still preserving much of the inertial range of the modeled turbulence resolved in the simulation.  In our models, we choose $\nu_3$ such that the mesh Reynolds number
\begin{equation} \label{E:remesh}
  \mathrm{Re}_\mathrm{mesh}\equiv\frac{u_\mathrm{max}}{\nu_3}
                                 \left(\frac{h}{\pi}\right)^5 \lesssim 1,
\end{equation}
where $u_\mathrm{max}$ is the absolute maximum of $u$ over the computational
domain.  This criterion states that dissipation at the Nyquist frequency should be comparable to or stronger than gas advection.  Notice that since all the adopted hyper-diffusive terms have the same coefficient $\nu_3$, the effective magnetic Prandtl number $\mathrm{Pr}_{M,\mathrm{eff}} \equiv \mu_0\nu_\mathrm{eff} / \eta_\mathrm{eff}$ in our simulations should be reasonably close to unity, where $\nu_\mathrm{eff}$ and $\eta_\mathrm{eff}$ are the effective viscosity and resistivity in the simulations, respectively.

  \subsection{Particle Dynamics} \label{SS:pd}

In this work, we consider particles of zero mass to study the effect of hydromagnetic turbulence on the orbital properties of planetesimals.  Although simple, this offers a good approximation for kilometer-sized planetesimals as they are large enough that gaseous drag force can be neglected, but small enough that type~I migration does not dominate.  For the approximation to be valid, the mass of a particle must lie between roughly $10^{14}$~g and $10^{26}$~g, corresponding to a size range of about 0.1--1000~km (OMM07).  Furthermore, if the action of hydromagnetic turbulence on the particles is separable from effects due to other interactions, our measurements can be applied to all stages of planet formation.

In our models, therefore, we consider particles moving under only the gravitational influence of the host star and that of the protoplanetary gas disk.  We ignore drag forces between the particles and the gas and the gravity of the particles.  Because the particles exert no force on the gas or themselves, no migration torques act. Given deterministic Keplerian shear flow and epicycle motions, deviations in particle trajectories due to turbulent fluctuations in the gas can easily be isolated.

Under these assumptions, the equations of motion for each particle become
\begin{mathletters} \label{E:eomp}
  \begin{eqnarray}
    \frac{\mathrm{d}\vec{x}_p}{\mathrm{d}t}
      &=& \vec{u}_p
          - \frac{3}{2}\Omega_K x_p\hat{\vec{y}},\label{E:eomp1}\\
    \frac{\mathrm{d}\vec{u}_p}{\mathrm{d}t}
      &=& \left(2\Omega_K u_{p,y}\hat{\vec{x}}
                - \frac{1}{2}\Omega_K u_{p,x}\hat{\vec{y}}\right)
          - \nabla\Phi.\label{E:eomp2}
  \end{eqnarray}
\end{mathletters}
The vector $\vec{x}_p$ is the position of the particle in the shearing box, while $\vec{u}_p$ is the velocity of the particle relative to the background shear flow.  The scalar variable $\Phi$ is the gravitational potential of the gas, which is the solution of the Poisson equation \citep{GL65}:
\begin{equation} \label{E:poisson}
  \left[\left(\partial_x -
              \frac{3}{2}\Omega_K t\partial_y\right)^2 +
        \partial_y^2 + \partial_z^2\right]\Phi = 4\pi G\rho,
\end{equation}
where $G$ is the gravitational constant.

Sheared, periodic boundary conditions \citep{HGB95} are adopted for the solution of  equation~\eqref{E:poisson}.  The system is strictly periodic in the $y$- and $z$-directions while the $x$-direction requires special treatment.  Since the $yz$-plane at any given $x$ moves with the Keplerian shear flow, the lower $x$-boundary plane should be imaged by the upper $x$-boundary plane shifted by $-3\Omega_K L_x\delta t / 2$ in the $y$-direction, where $\delta t$ is the time-step and $L_x$ is the $x$-dimension of the computational domain.  In mathematical notation, $\rho(x,y,z) = \rho(x + L_x, y - 3\Omega_K L_x\delta t / 2, z)$.  Rather than interpolating $\rho(x,y,z)$ in real space to obtain sheared periodicity, we use Fourier interpolation when solving equation~\eqref{E:poisson} in Fourier space \citep[see][Supplementary Information]{JO07}.\footnote{This technique was originally suggested by Colin McNally at \texttt{http://imp.mcmaster.ca/\%7Ecolinm/ism/rotfft.html}.}

The position $\vec{x}_p$ and velocity $\vec{u}_p$ of each particle is updated by solving the equations of motion~\eqref{E:eomp} simultaneously with the third-order Runge-Kutta steps for the MHD equations~\eqref{E:mhd_cont}--\eqref{E:mhd_ind}.  In addition to the Courant conditions set by the MHD equations, the time-step is limited by the absolute maximum of equation~\eqref{E:eomp1} such that no particles can cross more than half the zone size in one time-step.  We compute the gradient of the potential $\nabla\Phi$ on the grid after solving equation~\eqref{E:poisson} and then quadratically interpolate it to the position of each particle in the calculation of equation~\eqref{E:eomp2}.

  \subsection{Code Units and Scaling Relations} \label{SS:cusr}

We define the length and the time units as the vertical scale height $H$ and the orbital period $P = 2\pi / \Omega_K$, respectively, at the center of our local shearing box located at an arbitrary distance to the host star $R$.  Since vertical hydrostatic equilibrium of isothermal gas requires that $H = \sqrt{2} c_s / \Omega_K$, the speed of sound is fixed at $c_s = \pi\sqrt{2}$.  Note that this choice makes the system invariant with temperature.

We adopt two different mass units such that $\rho_0 = (4\pi G P^2)^{-1}$ and $\rho_0 = (G P^2)^{-1}$ for a low-mass and a high-mass disk, respectively, where $\rho_0$ is the uniform initial gas density.  For these two disk models, the Toomre $Q$ parameter for the gas is $Q_g = c_s\Omega_K / \pi G\Sigma = 63$ and 5.0, respectively, where $\Sigma = \sqrt{\pi}\rho_0 H$ is the column density.\footnote{Strictly speaking, this relation only holds for stratified disks.  Comparison between stratified and unstratified disk models will be made in a subsequent study.}  The gas disks in our models are gravitationally stable and thus we ignore gas self-gravity.  For convenience, we define a dimensionless parameter
\begin{equation}
  \xi \equiv 4\pi G\rho_0 P^2 = 4(2\pi)^{3/2} / Q_g
\end{equation}
as a measure of the strength of disk gravity.  For our low-mass and high-mass disks, $\xi = 1$ and $4\pi$, respectively.

In physical units, $\rho_0$ is given by the following scaling relation:
\begin{equation}
  \rho_0 = \left(1.2\times10^{-9}~\textrm{g cm}^{-3}\right)\xi
           \left(\frac{P}{\textrm{yr}}\right)^{-2} \nonumber\\
         = \left(1.2\times10^{-9}~\textrm{g cm}^{-3}\right)\xi
           \left(\frac{M_\star}{M_\sun}\right)
           \left(\frac{R}{\textrm{AU}}\right)^{-3},
\end{equation}
where $M_\star$ is the mass of the host star.  The corresponding column density is
\begin{equation}
  \Sigma = \left(1.5\times10^3\textrm{ g cm}^{-2}\right)\xi
           \left(\frac{c_s}{10^5\textrm{ cm s}^{-1}}\right)
           \left(\frac{P}{\textrm{yr}}\right)^{-1}.
\end{equation}
These scaling relations describe families of disk models to which our results apply.  In particular, at 1~AU around a solar-type star, the column density of our low-mass disk is roughly consistent with that of the classical minimum mass solar nebula \citep[MMSN;][]{cH81}.

Finally, the units of magnetic field and vector potential are $\mu_0^{1/2}\rho_0^{1/2}H P^{-1}$ and $\mu_0^{1/2}\rho_0^{1/2}H^2 P^{-1}$, respectively.  We arbitrarily set the permeability $\mu_0 = 1$.  The magnetic energy density associated with $B = 1$ is then $1 / (4\pi^2)$ of the initial pressure $p_0 = c_s^2\rho_0$.

  \subsection{Initial and Boundary Conditions} \label{SS:ibc}

The initial conditions for our models are the following.  The gas density is uniform ($\rho = \rho_0$) while the magnetic vector potential is set to zero ($\vec{A} = 0$).  An external vertical magnetic field is imposed, ranging from
$B_\mathrm{ext} = 0.01$ to 0.64.  The corresponding plasma $\beta \equiv 2\mu_0\rho c_s^2 / B^2$ ranges from $3.9\times10^5$ to 96.  Gaussian noise in gas velocity of amplitude $10^{-3}$ is imposed to seed the MRI.

As noted in \S\ref{SS:pd}, the boundary conditions for all dynamical variables are sheared periodic and we find values for the ghost zones using Fourier interpolation.  Our fiducial model has a computational domain of $2\times2\times2H$, but we also study domains with sizes up to $8\times8\times2H$.  The highest resolution we use is 64 grid points per disk scale height $H$.

The coefficient $\nu_3$ of the hyper-diffusive terms discussed in \S\ref{SS:mhd} needs to be fine-tuned such that the mesh Reynolds number $\mathrm{Re}_\mathrm{mesh}$ is as close to unity as possible during the course of the simulation (eq.~[\ref{E:remesh}]).  We adopt an iterative approach to determine the optimal value of $\nu_3$, using $\mathrm{Re}_\mathrm{mesh} \sim 1$.  For the case of $B_\mathrm{ext} = 0.08$ ($\beta_\mathrm{ext} = 6.2\times10^3$) with a $2\times2\times2H$ box at a resolution of 64~points per scale height, we choose $\nu_3 = 2.9\times10^{-11}$ for which $u_\mathrm{max} \simeq 6.0\pm2.7$ at saturation level, where the deviation of $u_\mathrm{max}$ is given by $3\sigma$ in its time variation.

We uniformly distribute 32$^3$ particles in the entire computational domain.  We do not allow them to move until time $t = t_0$ after which the hydromagnetic turbulence has saturated and approached a statistically steady state.  For the case of $B_\mathrm{ext} = 0.08$ ($\beta_\mathrm{ext} = 6.2\times10^3$), we choose $t_0 = 20P$ (see \S\ref{SS:conv}).  Then the particles are set to initially move relative to the background shear flow such that they have an initial eccentricity of $e_0$ and start at the apogee of their orbits, i.e.,
\begin{equation} \label{E:icp}
  \vec{u}_{p,0} =
  -\frac{1}{2}H\Omega_K\left(\frac{e_0}{H / R}\right)\hat{\vec{y}}
\end{equation}
(see Appendix~\ref{S:vpao}).  We wrap a particle around when it moves beyond any of the six boundary planes.

We remark that each model presented in the following section is just one realization of the stochastic nature of the turbulence, corresponding to one set of initial velocity perturbations of the gas.  The similarity in particle orbital evolutions found across several models are due to closeness of the random number sequences used to generate the velocity perturbations and thus similar initial conditions for the gas.

\section{SIMULATION RESULTS} \label{S:sr}

\subsection{Convergence of Turbulence Properties} \label{SS:conv}

\subsubsection{Grid Resolution}
\begin{figure}[!htbp]
\begin{center}
\epsscale{0.8}
\plotone{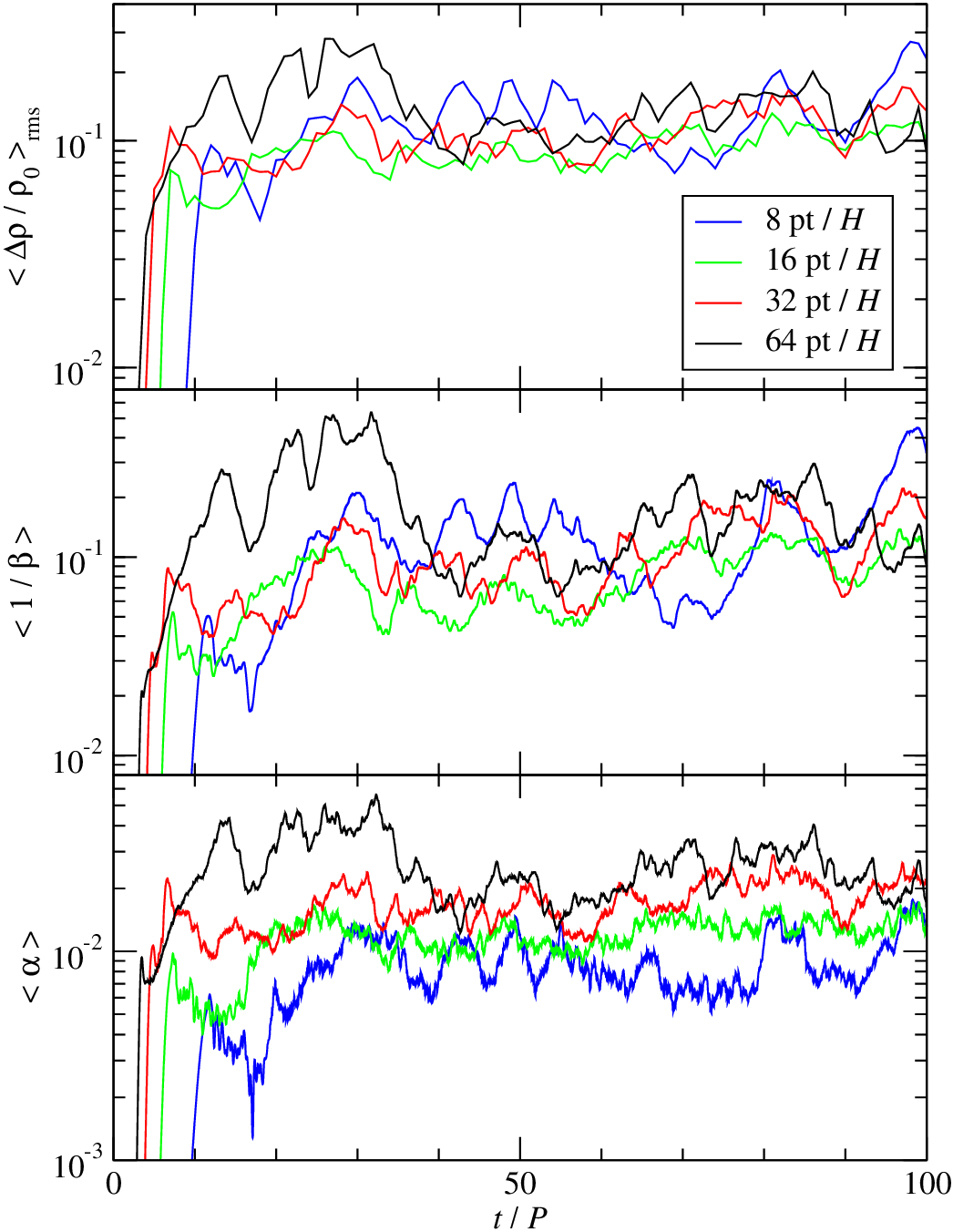}
\caption{Density perturbation $\Delta\rho / \rho_0$, inverse plasma $\beta$, and $\alpha$-parameter as a function of time $t$ (in units of orbital period $P$) for a $2\times2\times2H$ local shearing box under an external vertical magnetic field of $B_\mathrm{ext} = 0.08$ ($\beta_\mathrm{ext} = 6.2\times10^3$).  All properties are volume-averaged over the whole computational domain with the root-mean-square value of $\Delta\rho / \rho_0$ being given.  Results are shown for resolutions up to 64~points per scale height $H$.}
\label{Fi:mrit}
\end{center}
\end{figure}

We first present a study of the convergence with increasing numerical resolution of the properties of the turbulence important to our work.  We work on a $2\times2\times2H$ grid for this study to reach maximum resolution.  Figure~\ref{Fi:mrit} plots density perturbation $\Delta\rho / \rho_0$, inverse plasma $\beta$, and the \citet{SS73} $\alpha$-parameter as a function of time $t$ for disks with an external magnetic field of $B_\mathrm{ext} = 0.08$ ($\beta_\mathrm{ext} = 6.2\times10^3$) at resolutions up to 64~grid points per disk scale height $H$, where $\Delta\rho \equiv \rho - \rho_0$.  The density perturbation $\Delta\rho / \rho_0$ shown is the root-mean-square value over the computational domain while the inverse plasma  $\beta$ shown is the volume-averaged value.  The $\alpha$-parameter is calculated from the combined effects due to the Reynolds and Maxwell shear stresses \citep[e.g.,][]{aB98}:
\begin{equation}
  \alpha = \frac{\sqrt{2}}{3}
           \frac{\langle\rho u_x u_y - B_x B_y / \mu_0\rangle}{\rho_0 c_s^2},
\end{equation}
where the bracket $\langle\rangle$ denotes the volume average over the entire computational domain.  As shown in Figure~\ref{Fi:mrit}, the MRI saturates and remains roughly steady after about $t = 20P$. After saturation, all three properties exhibit only small changes with increasing resolution, aside from a slight trend of increasing $\alpha$.  The properties of the saturated turbulence appear to converge to a nonzero level, as opposed to disks without net magnetic flux \citep[e.g.,][]{FP07a}.

\subsubsection{Box Size}

\begin{figure}[!htbp]
\begin{center}
\plotone{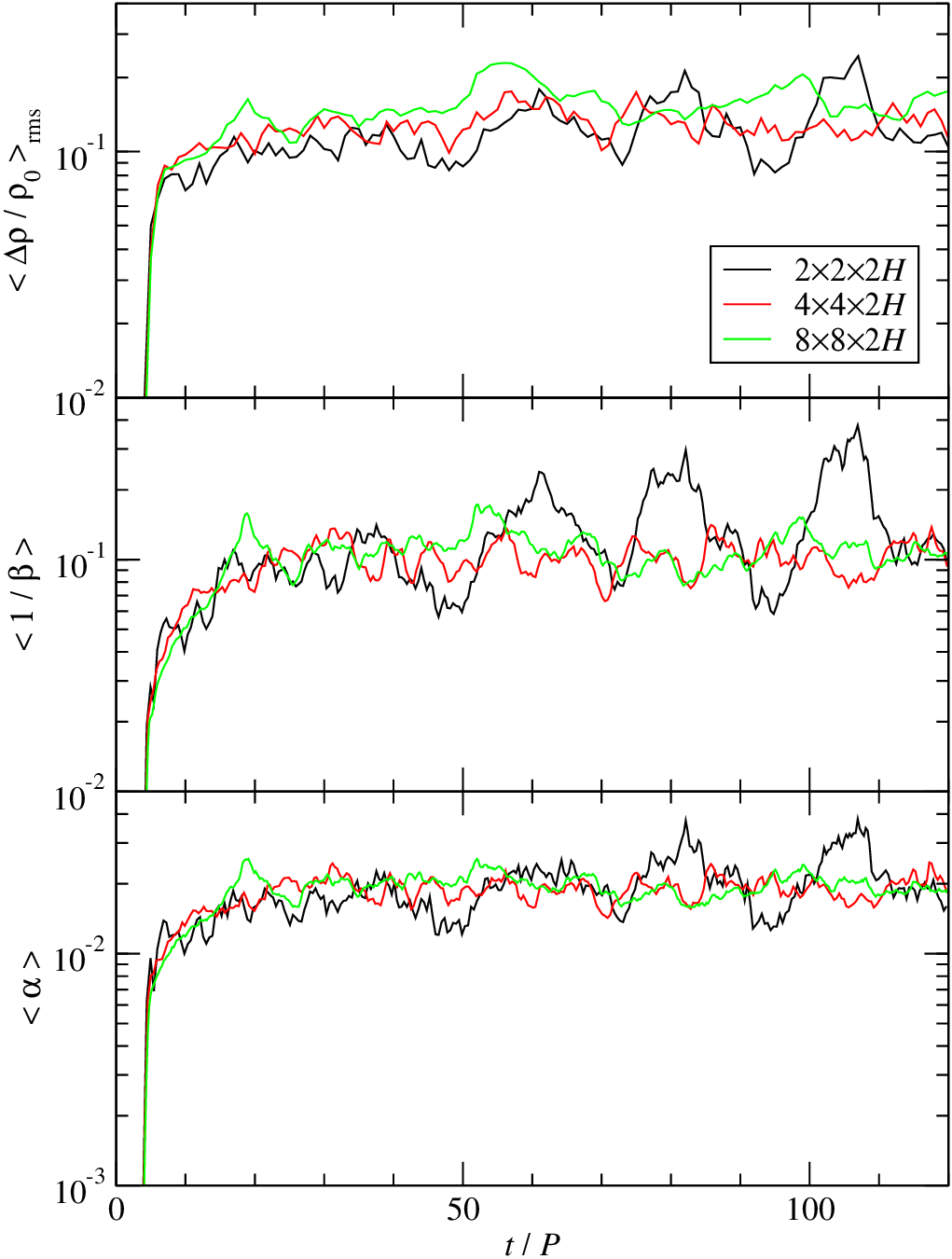}
\caption{Density perturbation $\Delta\rho / \rho_0$, inverse plasma $\beta$, and $\alpha$-parameter as a function of time $t$ for three different box sizes at a resolution of 32~points per scale height $H$.  An external vertical magnetic field of $B_\mathrm{ext} = 0.08$ ($\beta_\mathrm{ext} = 6.2\times10^3$) is imposed.  Properties are volume-averaged over the whole computational domain and the root-mean-square value for $\Delta\rho / \rho_0$ is shown.}
\label{Fi:mrit_box}
\end{center}
\end{figure}

We also examine how the same averaged quantities depend on the size of our local simulation box by studying three runs done with increasing horizontal size, up to $8\times8\times2H$, at our medium resolution of 32 points per scale height.  \citet{JYK09} ran high-resolution (with $\sim$137~points per scale hight), unstratified models without mean field and found roughly linear growth in the effective viscosity $\alpha$ with box size.  In Figure~\ref{Fi:mrit_box} we show that the plasma $\beta$ and the effective $\alpha$-parameter appear almost independent of box size in our models, aside from a trend toward reduced temporal fluctuations when averaged over larger boxes.  The root-mean-square density perturbation $\Delta\rho / \rho_0$ shows a weak trend towards increasing at larger box size, though, with the average over the time period $20 < t/P < 120$ increasing by 26\% from the smallest to the largest box, a scale change of a factor of four.  This probably occurs because of the inclusion of larger-scale instability modes in the larger boxes; it is not a particularly dramatic effect, though, because the smallest box size that we study already captures the fastest growing modes.  Nevertheless, this small effect seems to strongly affect particle orbital properties, as we will discuss below.

  \subsection{Vertical Net Flux Dependence} \label{SS:vnfdp}

\begin{figure}[!htbp]
\begin{center}
\epsscale{0.75}
\plotone{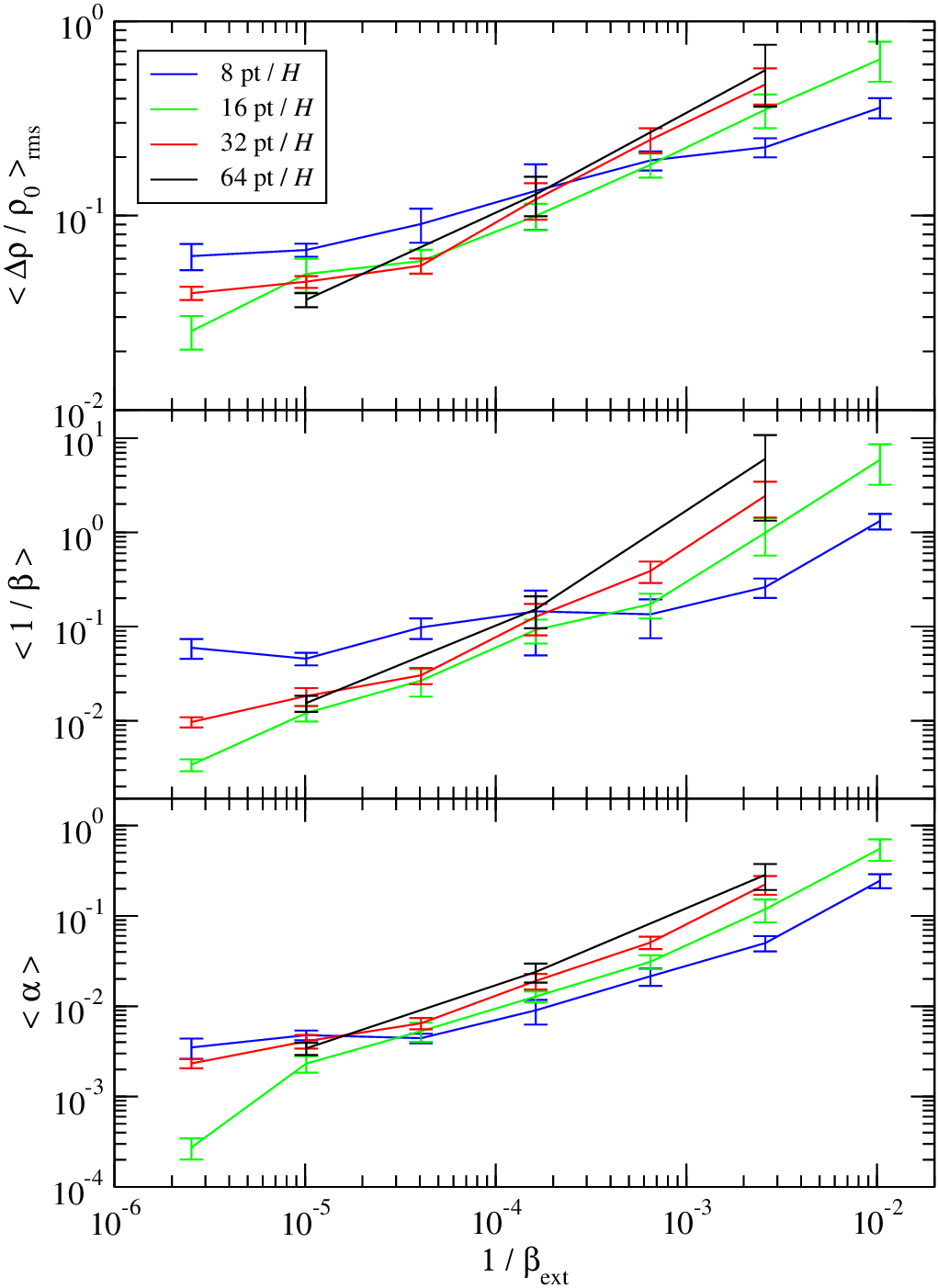}
\epsscale{0.8}
\caption{Density perturbation $\Delta\rho / \rho_0$, inverse plasma $\beta$, and $\alpha$-parameter as a function of external magnetic field in terms of inverse plasma $\beta$ for a $2\times2\times2H$ box.  All properties are volume-averaged over the whole computational domain as well as time-averaged over an interval of at least 20 orbital periods after saturation of the MRI.  Results are shown for resolutions up to 64~points per scale height $H$.  The error bars denote $1\sigma$ in time variation around the volume-averaged properties.}
\label{Fi:mrib}
\end{center}
\end{figure}

We next turn to the effect of varying external vertical magnetic field.  Figure~\ref{Fi:mrib} plots the same properties of the saturated turbulence as a function of the external vertical magnetic field, represented by the inverse plasma $\beta$, for our smallest box.  They are volume averaged as described above, and then time-averaged over a period of at least $20P$ after saturation.  Also included in the figure are the time variation of these properties, as indicated by the error bars.  We confirm the general trend of increasing turbulence activity with increasing uniform vertical field \citep[e.g.,][]{HGB95,SI04,JKM06}.  Numerical convergence can be seen over the range of the field strengths we have explored.  In addition to the turbulent transport often discussed in the literature, we also report the dependency between density perturbation and external field in Figure~\ref{Fi:mrib}, which may be more relevant to the orbital dynamics of particles moving in these disks.  We emphasize that by varying the net vertical magnetic flux through a disk, a wide range of turbulent viscosity values can be obtained, as suggested by numerous previous works as well as Figure~\ref{Fi:mrib}.

To best represent typical protoplanetary accretion disks, we adopt a fiducial disk model with $B_\mathrm{ext} = 0.08$ ($\beta_\mathrm{ext} = 6.2\times10^3$), which we run at a resolution of 64 points per scale height on a $2\times2\times2H$ grid.  As shown in Figures~\ref{Fi:mrit} and \ref{Fi:mrib}, this model gives a turbulent accretion of $\alpha\sim10^{-2}$, which is consistent with current estimates for disks around typical T~Tauri stars \citep[e.g.,][]{HC98,HD06}.  Note that in our fiducial model, the root-mean-square density perturbation of the gas is on the order of $\sim$10\%.

  \subsection{Motion of a Single Particle} \label{SS:msp}

\begin{figure}[!htbp]
\begin{center}
\plotone{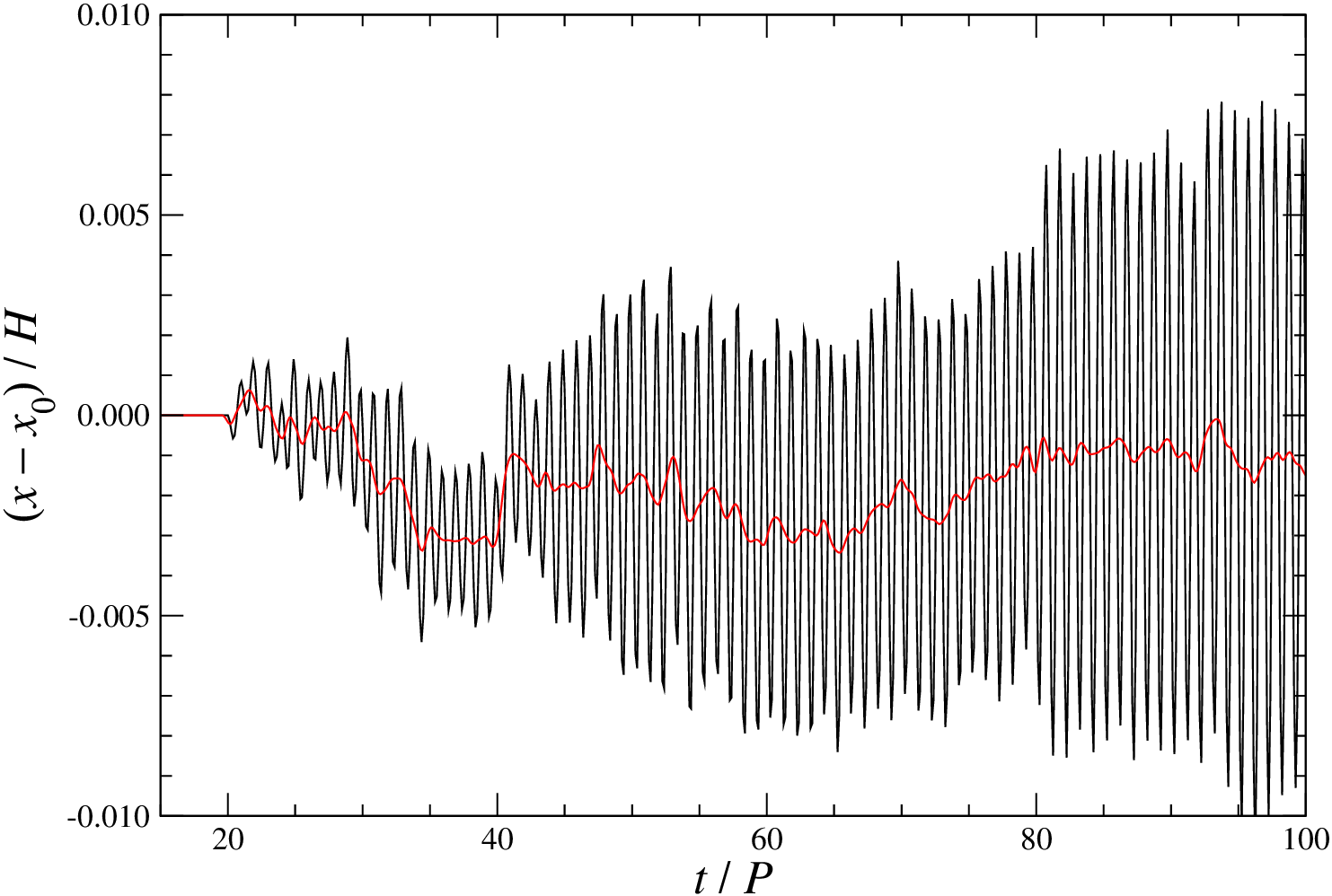}
\caption{Radial motion of one representative particle with initial eccentricity $e_0 = 0$ moving in the low-mass disk version of our fiducial model.  The \emph{black} line shows its radial displacement $x - x_0$ from initial position $x_0$ as a function of time $t$.  The \emph{red} line shows the corresponding radial drift, defined as the running average over one epicycle, i.e., one orbital period $P$.}
\label{Fi:rd}
\end{center}
\end{figure}

Using our fiducial model in the context of a protoplanetary disk, we now study the orbital dynamics of zero-mass particles moving in this turbulent environment.  As demonstrated in Figure~\ref{Fi:rd}, a particle moving under gravity of the turbulent gas undergoes epicycle motion horizontally as well as continuous change in its mean radius.  We define the radial drift of each particle as $\Delta x \equiv \bar{x} - x_0$, where $\bar{x}$ is the mean radial position over one orbital period (as exemplified by the red line in Fig.~\ref{Fi:rd}) and $x_0$ is the initial radial position.  Given that $x \ll R$, the eccentricity of each particle can be approximated by $e \approx \left(x_\mathrm{max} - x_\mathrm{min}\right) / 2R$, where $x_\mathrm{max}$ and $x_\mathrm{min}$ are the maximum and the minimum radial positions in one epicycle, respectively.  With these two quantities, we can measure the orbital migration and eccentricity change of planetesimals induced by hydromagnetic turbulence.

\begin{figure}[!htbp]
\begin{center}
\plotone{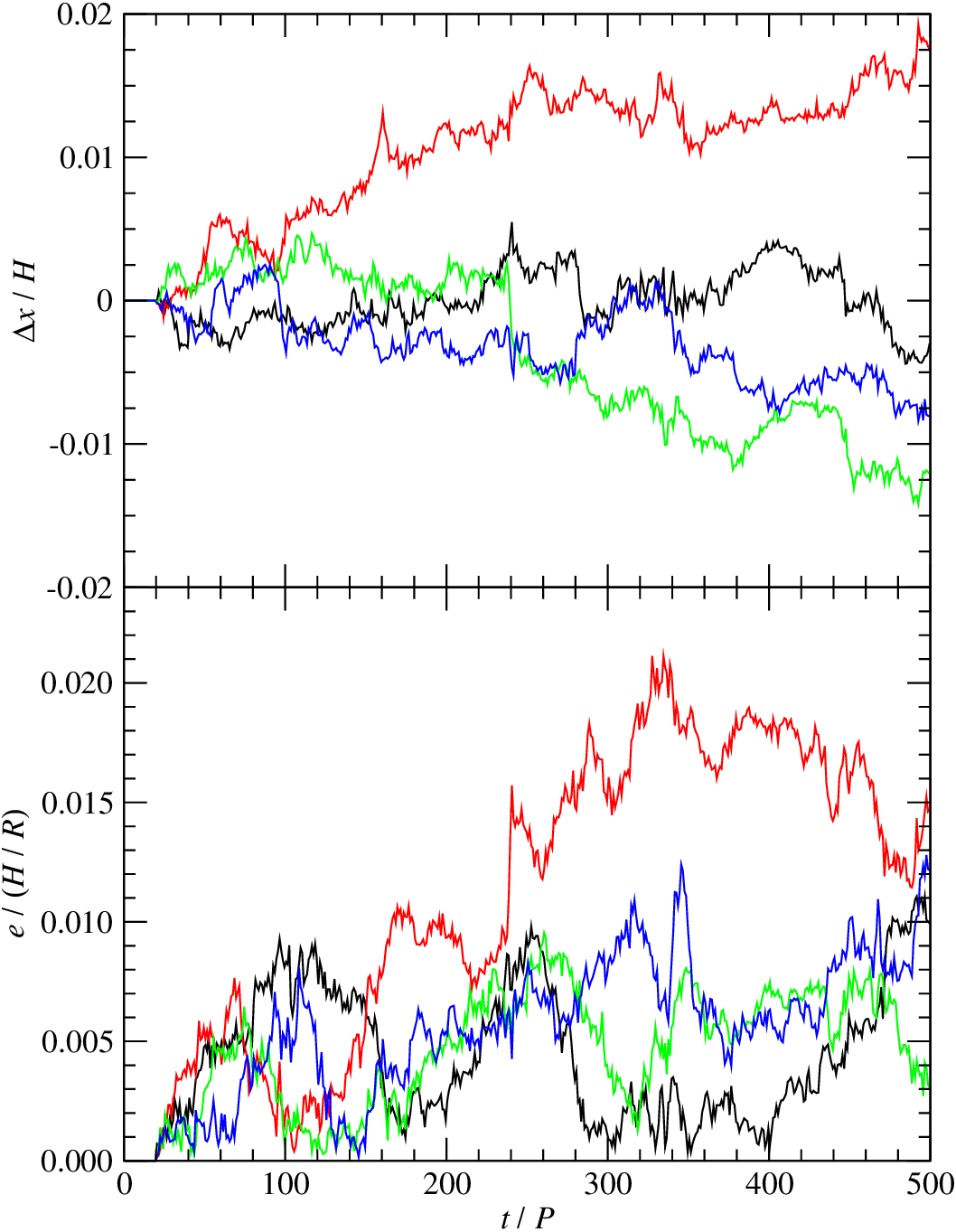}
\caption{Time evolution of radial drift $\Delta x$ (\emph{top} panel) and eccentricity $e$ (\emph{bottom} panel) for four randomly selected particles with initial eccentricity $e_0 = 0$ moving in the low-mass disk version of our fiducial model.  The eccentricity is in terms of $H / R$, the ratio of one disk scale height to the distance of the shearing box to the host star.}
\label{Fi:dx-e}
\end{center}
\end{figure}

Figure~\ref{Fi:dx-e} shows the change of radial drift and eccentricity with time for four randomly selected particles.  It is evidently a stochastic process and the final outcome can be quite different with slightly different initial conditions.  Nevertheless, statistical methods can be employed to quantify the process.  We discuss the statistical evolution of these orbital properties in the following subsections.
  
  \subsection{Radial Drift} \label{SS:rd}

\begin{figure}[!htbp]
\begin{center}
\plotone{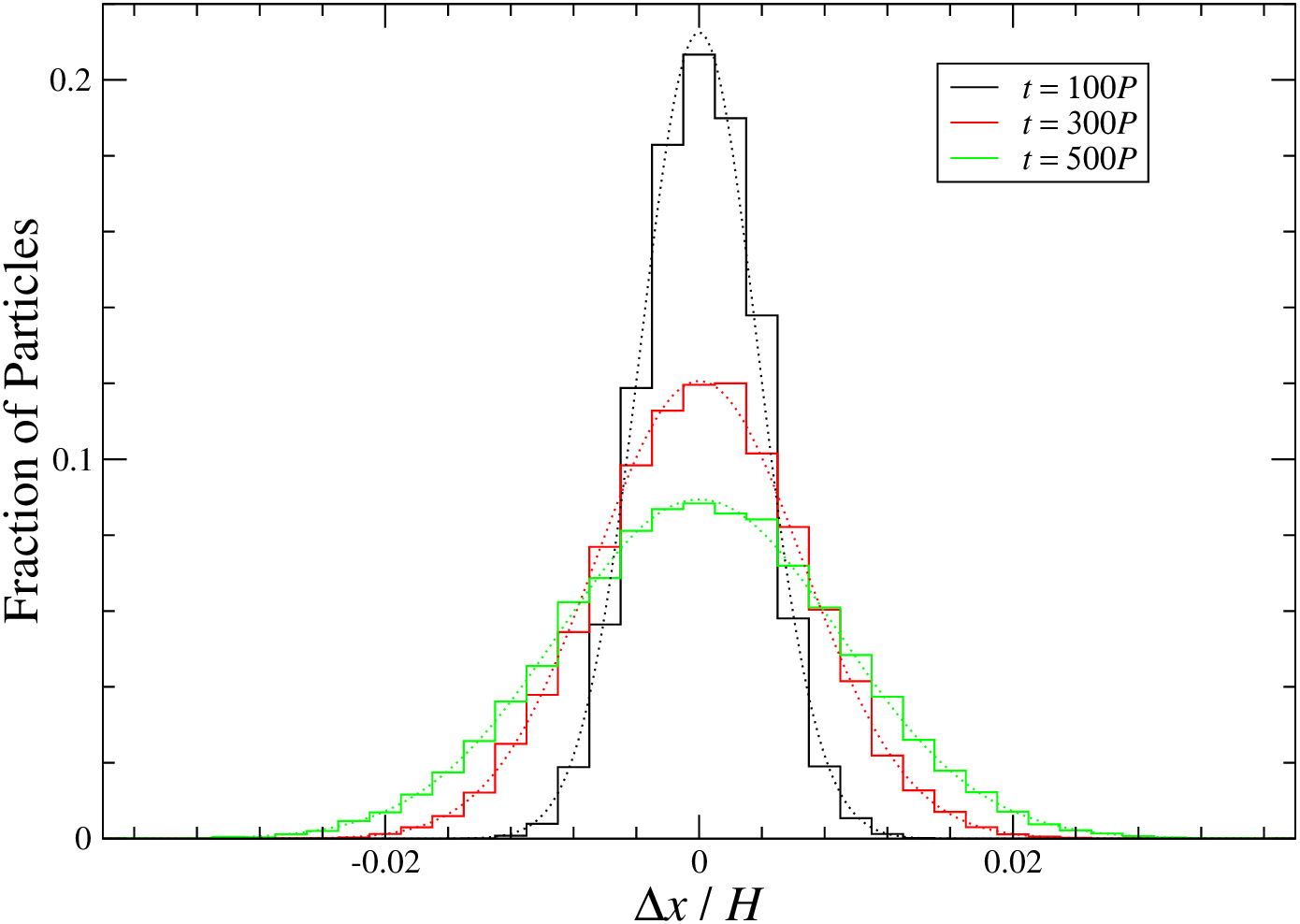}
\caption{Distribution of particle radial drifts $\Delta x$ at three different times in our fiducial model.  These particles have initial eccentricity $e_0 = 0$ and evolve in a low-mass disk.  The dotted lines are best fits in the form of normal distributions.}
\label{Fi:dx_dist}
\end{center}
\end{figure}

Histograms of the distribution of radial drifts at three different times in the low-mass disk version of our fiducial model are plotted in Figure~\ref{Fi:dx_dist}.  The distribution of particles in radial drift resembles a normal distribution with its center located at approximately zero.  This is not surprising since there is no preferred direction locally for the turbulence to generate a net torque.  More interestingly, the width of the distribution increases with time.  Although the hydromagnetic turbulence has no net effect on the orbital radius of the particles, it becomes more and more likely for any single particle to drift away from its original orbit as time increases.  This could help a subset of particles to survive type~I migration, as suggested by JGM06 and \citet{AB09}.

\begin{figure}[!htbp]
\begin{center}
\plotone{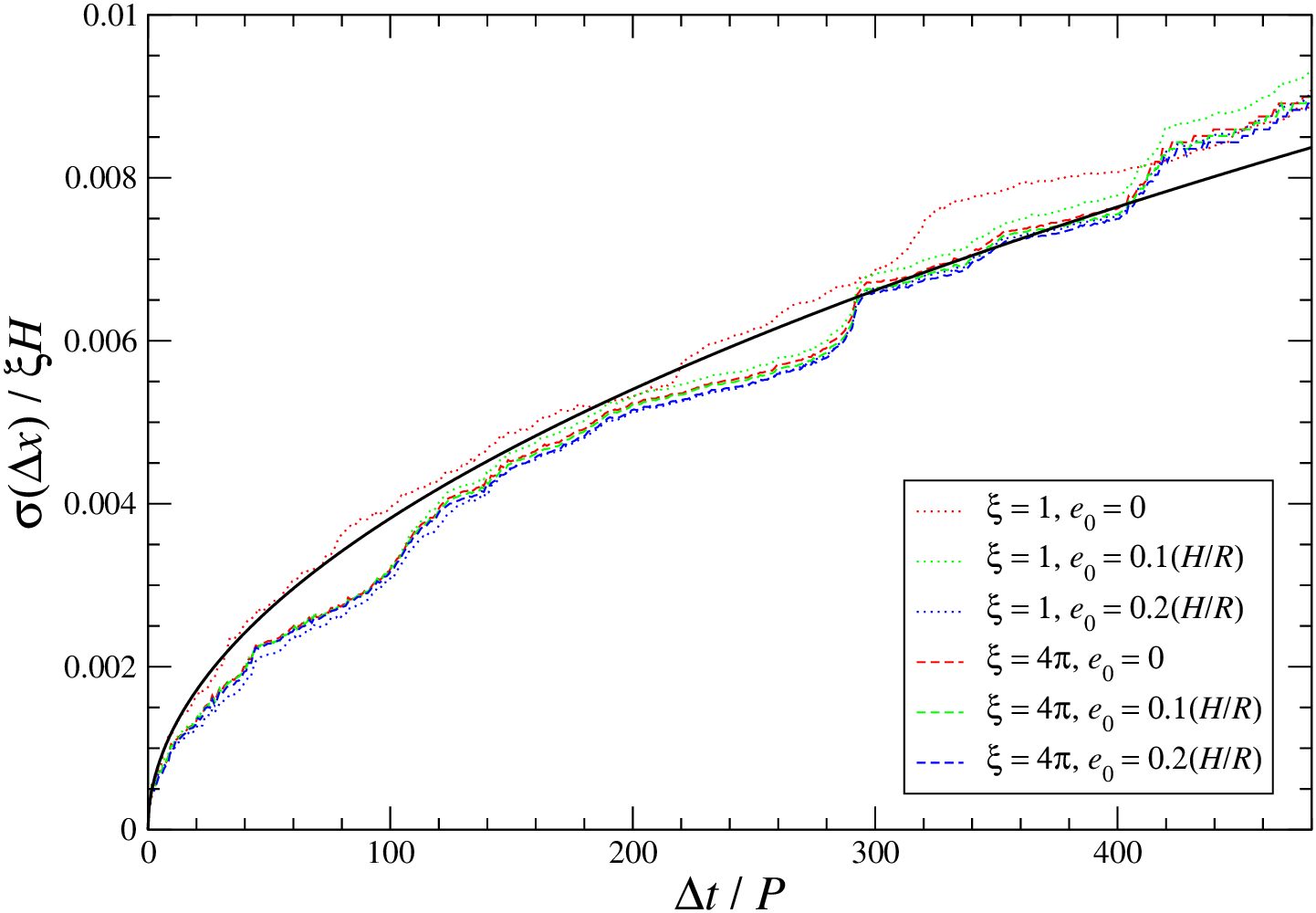}
\caption{Standard deviation of radial drift $\sigma(\Delta x)$ as a function of elapsed time $\Delta t$ in our fiducial model.  The \emph{dotted} lines are obtained from low-mass disks ($\xi = 1$) while the \emph{dashed} lines are from high-mass disks ($\xi = 4\pi$).  Particles with initial eccentricities $e_0 = 0$, $0.1(H/R)$, and $0.2(H/R)$ are denoted by \emph{red}, \emph{green}, and \emph{blue} lines, respectively.  The \emph{black solid} line is the best fit to all six curves.}
\label{Fi:dxt}
\end{center}
\end{figure}

Figure~\ref{Fi:dxt} shows the standard deviation of radial drift $\sigma(\Delta x)$ as a function of elapsed time $\Delta t \equiv t - t_0$ for particles with different initial eccentricity $e_0$ moving in different strengths of disk gravity $\xi$ in our fiducial model.  The standard deviation steadily increases with time, with little difference between particles of different initial eccentricities.  Power-law fitting results in time indices of about 0.52--0.58, just slightly larger than 1/2.  This confirms the proposition that gravitational influence of the hydromagnetic turbulence makes particles undergo random walks (LSA04; \citealt{NP04}; N05) and the resulting orbital evolution can be described as a diffusion process \citep[JGM06; OIM07;][]{AB09,RP09}.

We measure the dependence of radial drift on disk gravity, as quantified by the dimensionless parameter $\xi$, by comparing the results from our low-mass and high-mass disk models (see \S\ref{SS:cusr}).  As demonstrated in Figure~\ref{Fi:dxt}, radial drifts in these models roughly coincide after being scaled with $\xi$, indicating a dependence close to linear.  By assuming that $\sigma(\Delta x)$ scales with $\xi\Delta t^{1/2}$, our best fit to the results shown in Figure~\ref{Fi:dxt} is
\begin{equation} \label{E:dx}
  \sigma(\Delta x) = (3.8\pm0.4)\times10^{-4}\,\xi H
                     \left(\frac{\Delta t}{P}\right)^{1/2}.
\end{equation}

\begin{figure}[!htbp]
\begin{center}
\plotone{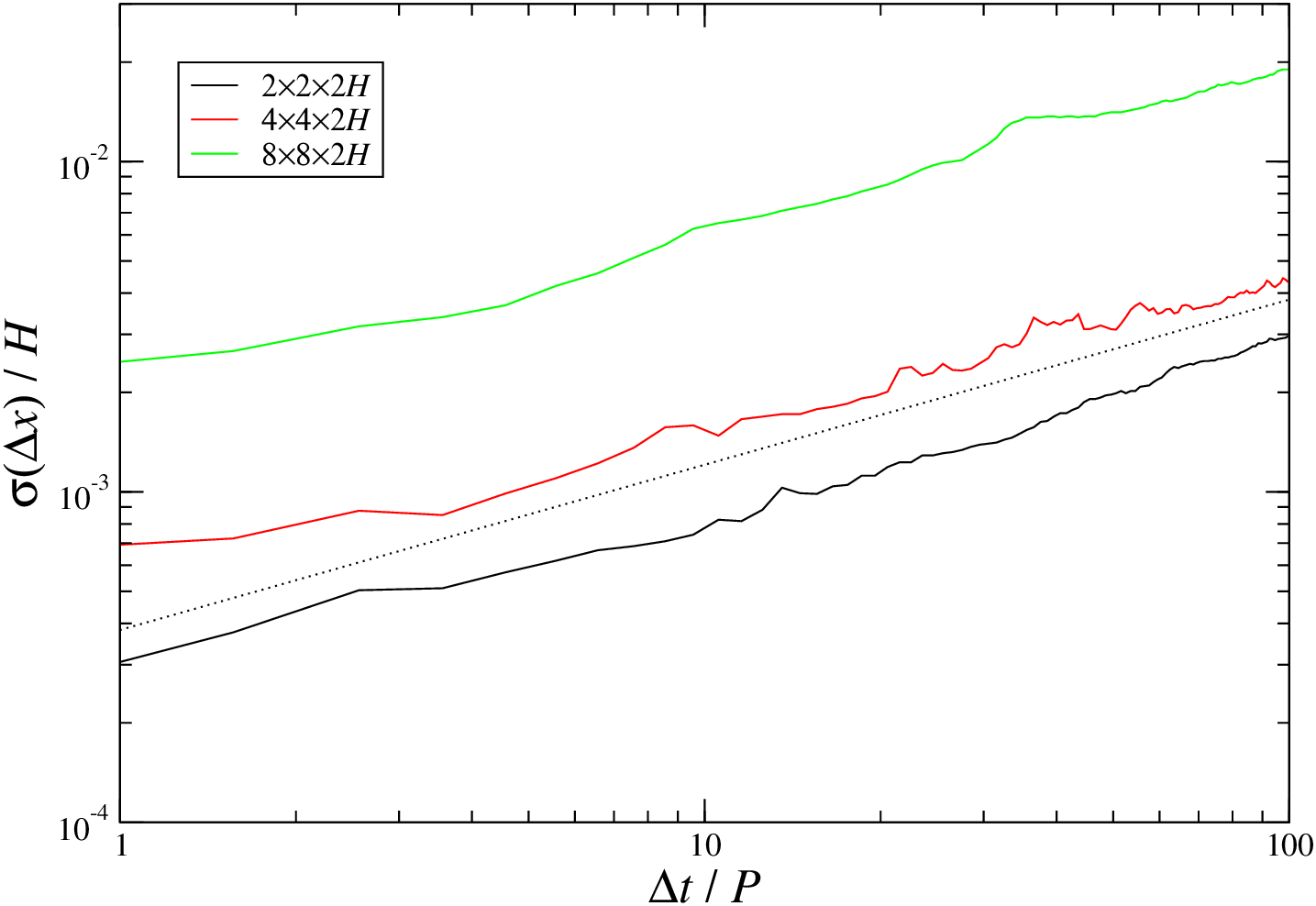}
\caption{Standard deviation of radial drift $\sigma(\Delta x)$ as a function of elapsed time $\Delta t$ for three different box sizes at a resolution of 32~points per scale height $H$ (\emph{solid} lines), where only the low-mass disk model ($\xi = 1$) and particles with zero initial eccentricity are considered.  For comparison, the straight \emph{dotted} line gives the best fit to the high-resolution model shown in Figure~\ref{Fi:dxt} (eq.~[\ref{E:dx}]).}
\label{Fi:dxt_box}
\end{center}
\end{figure}

The size of the computational domain of a local shearing box does have a rather substantial effect on the magnitude of the random walk, and thus the diffusion derived from this model, as shown in Figure~\ref{Fi:dxt_box}.  Increasing the horizontal size of the box by a factor of four results in almost an order of magnitude increase in the standard deviation of the radial drift at each time.  In contrast to gas properties discussed in \S\ref{SS:conv}, we have not seen convergence of the magnitude of the orbital random walk with box size in our study.

Our largest box shows an amplitude of the random walk roughly a factor of three smaller than seen in the global model described by N05.  We make this estimate by examining his Figure~4.  This shows the semi-major axis for six zero-mass particles at different radii $R$ over a period of time of roughly 100~orbits in the relevant region of 2--3 times the inner edge of his grid, at $R = R_1$, in a disk with $H \simeq 0.1 R$.  The standard deviation of $\Delta x / \xi H$ of the particles at $\Delta t \simeq 90P$ is roughly 0.06, where $P$ is the orbital period at the respective initial radii of the particles and the relevant disk-gravity parameter $\xi \simeq 0.9\left(R / R_1\right)^2$ (Richard~P.\ Nelson 2009, private communication).  By comparison, the result for our $8\times8\times2H$ box shown in Figure~\ref{Fi:dxt_box} gives $\sigma(\Delta x) / H \simeq 0.02$ at $\Delta t \simeq 90P$.

On the other hand, OIM07 found results roughly consistent with our fiducial disk model.  These authors used orbital integrations of particles influenced by torques given by a heuristic, stochastic formula for hydromagnetic turbulence, and the formula was suggested by LSA04 based on zero net flux, global disk, MHD simulations.  The disk models OIM07 studied were about 10--100 times less massive than the MMSN, but they reported their scaling with varying disk mass.  By extrapolating their results to values appropriate for our low-mass disk model with $\xi = 1$, and considering their fiducial magnitudes for the stochastic torques, we find that our measured spread of radial drift (eq.~[\ref{E:dx}]) is roughly consistent with theirs at 1~AU.  As discussed by these authors, large uncertainty in thier results might be involved, mostly due to the uncertainty in the magnitude of the stochastic torques and their neglecting the power given by LSA04 in the $m = 1$ mode (where the integer $m$ represents the Fourier decomposition of density structure in azimuthal angle rather than the spiral mode used in density wave theories).

  \subsection{Eccentricity} \label{SS:ecc}

\begin{figure}[!htbp]
\begin{center}
\epsscale{0.7}
\plotone{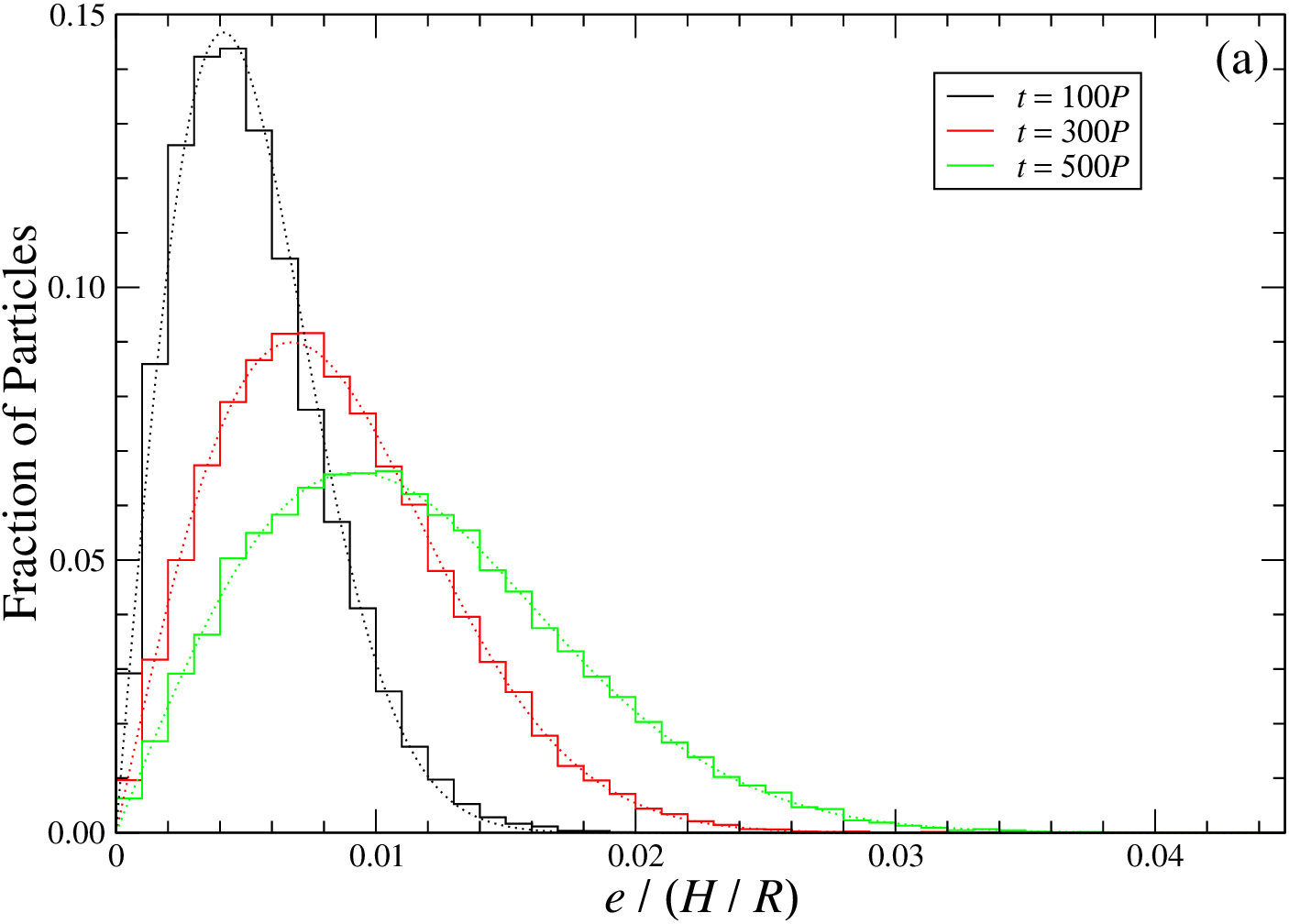} \\[.6in]
\plotone{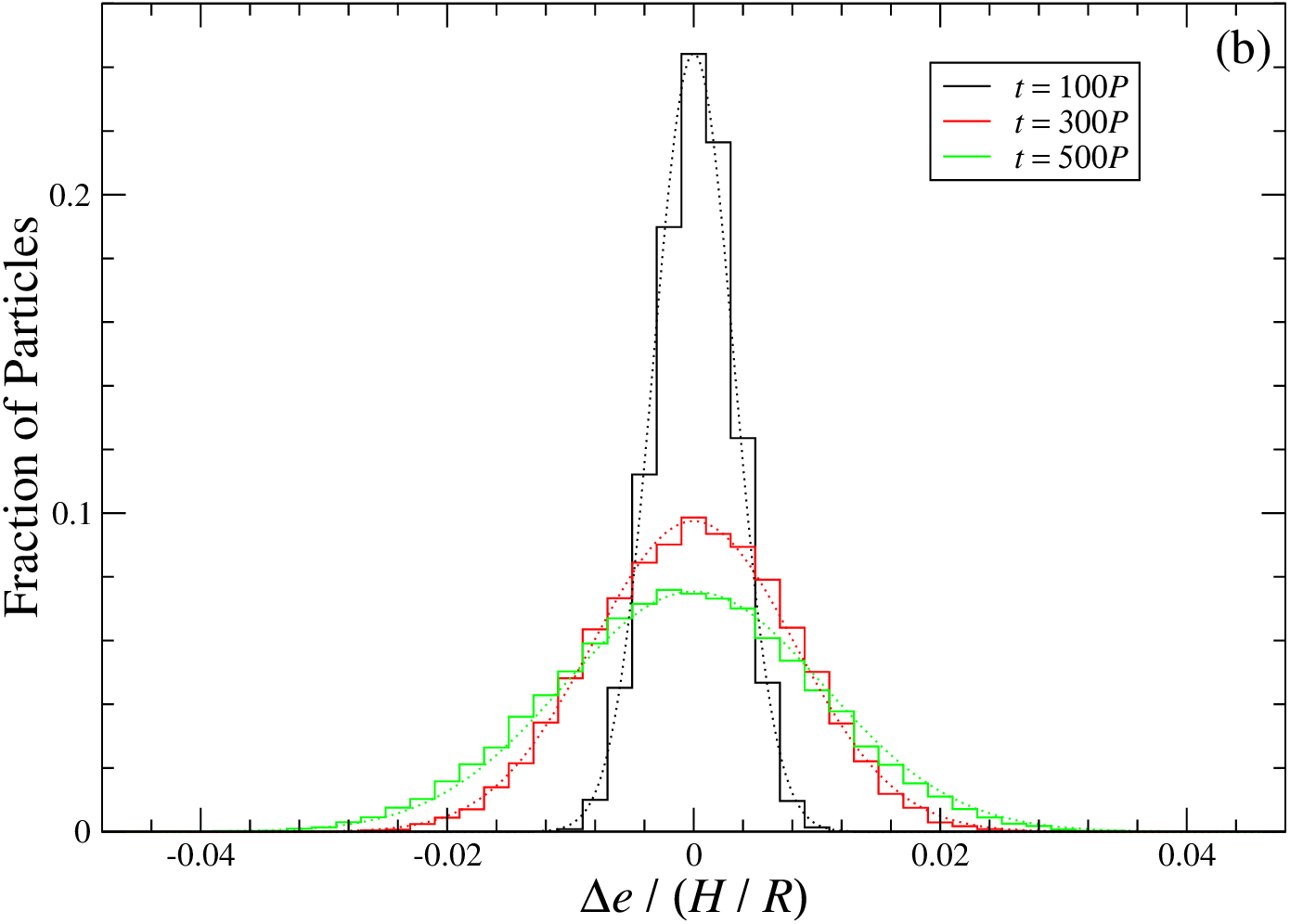}
\epsscale{0.8}
\caption{Distribution of particles in (a)~eccentricity $e$ for particles with initial eccentricity $e_0 = 0$ and (b)~eccentricity deviation $\Delta e = e - e_0$ for particles with $e_0 = 0.1(H / R)$ at three different times in our fiducial model.  These particles move in a low-mass disk.  The dotted lines are best fits in the form of (a)~Rayleigh and (b)~normal distributions.}
\label{Fi:de_dist}
\end{center}
\end{figure}

Figure~\ref{Fi:de_dist}a shows the histograms of eccentricity at three different times for particles with initial eccentricity $e_0 = 0$ moving in the low-mass version of our fiducial disk model.\footnote{We remark that the mean eccentricity of particles with initial eccentricity $e_0 = 0$ is about $0.012(H/R)$ and $0.16(H/R)$ at $t = 500P$ for the low-mass and the high-mass disk models, respectively.  These eccentricities correspond to epicycle motions covering about 1.5 and 20~grid zones in the radial direction, enough to resolve the gravitational forces experienced by the particles.}  The distributions appear close to a Rayleigh distribution:
\begin{equation} \label{E:rayleigh}
  f(e) = \left(\frac{e}{\sigma^2}\right)\exp\left(-\frac{e^2}{2\sigma^2}\right),
\end{equation}
where $\sigma$ is a constant proportional to both the peak and the width of the distribution.  The standard deviation of this distribution is given by $\sigma\sqrt{(4 - \pi) / 2}$.  As shown in Figure~\ref{Fi:de_dist}a, both the eccentricity at the peak value and the width of the distribution increase with time, so it at first appears that eccentricity is excited by hydromagnetic turbulence.  Such eccentricity growth was also reported by N05 and OIM07. 

However, we find a different distribution for particles with nonzero initial eccentricity in the same model.  Figure~\ref{Fi:de_dist}b shows histograms of eccentricity deviation $\Delta e = e - e_0$ at three different times for particles with initial eccentricity $e_0 = 0.1(H/R)$.  They are similar to a normal distribution with a constant mean, implying that the average eccentricity of these particles remains constant at the initial eccentricity (see also OIM07).  The width of the distribution does increase with time.  Therefore, hydromagnetic turbulence does not just excite the eccentricity of particles; it can also act to damp the existing eccentricity of some particles.  The distribution found for particles with $e_0 = 0$ is just a special case of this general behavior: since eccentricity is a positive definite quantity and the initial eccentricity is zero, it is no surprise that a normal distribution for the eccentricity deviation, $f(\Delta e) = \exp\left(-\Delta e^2 / 2\sigma^2\right) / \sqrt{2\pi}\sigma$, manifests as a Rayleigh distribution for the eccentricity itself (eq.~[\ref{E:rayleigh}]).

\begin{figure}[!tbp]
\begin{center}
\plotone{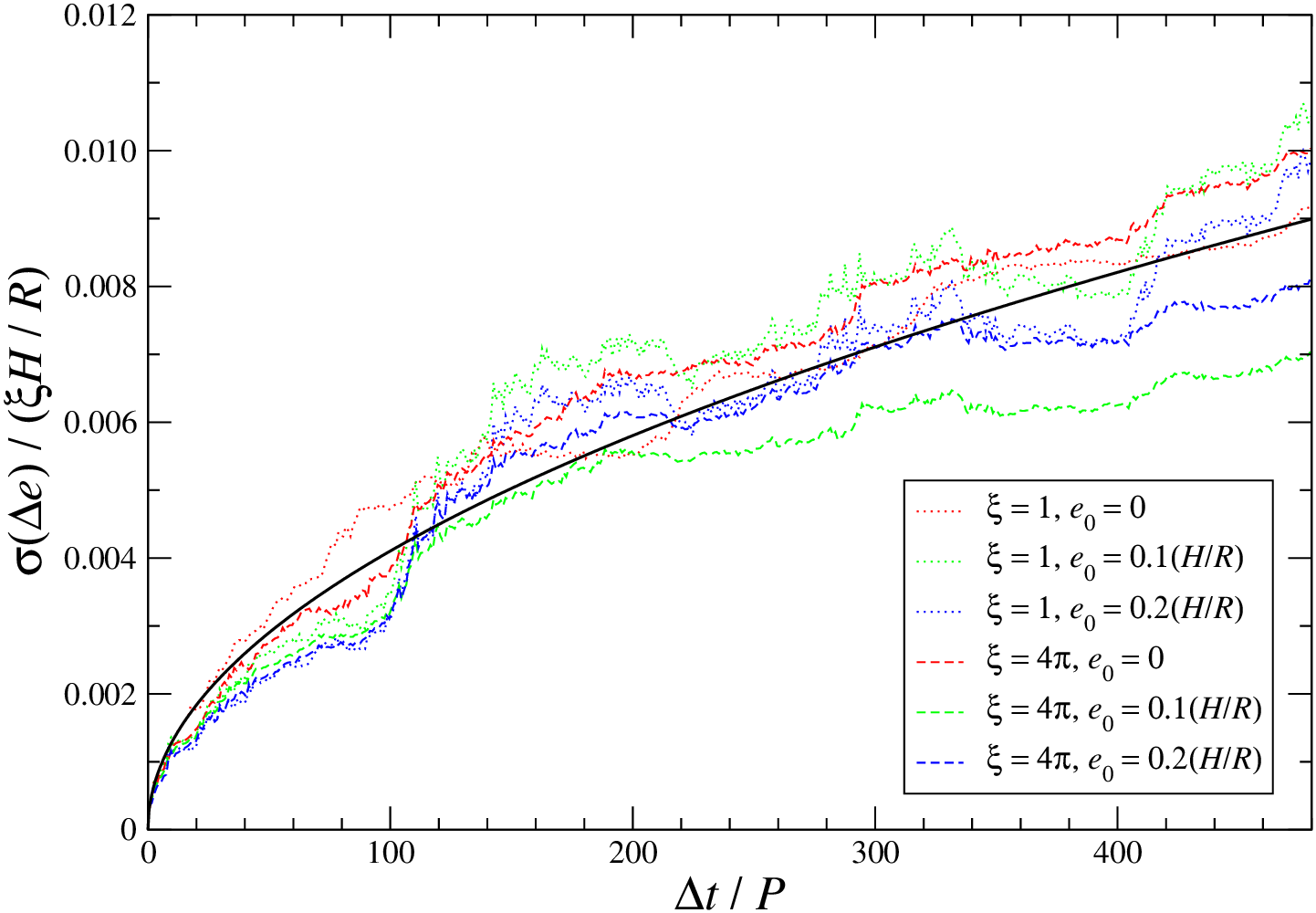}
\caption{Standard deviation of eccentricity deviation $\sigma(\Delta e)$ as a function of elapsed time $\Delta t$ in our fiducial model.  The line styles and colors are the same as in Figure~\ref{Fi:dxt}.  The standard deviation measured for particles with initial eccentricity $e_0 = 0$ has been multiplied by a factor of $\sqrt{2 / (4 - \pi)}$ to be consistent with a normal distribution (see \S\ref{SS:ecc}).}
\label{Fi:det}
\end{center}
\end{figure}

Figure~\ref{Fi:det} shows the standard deviation of eccentricity deviation $\sigma(\Delta e)$ as a function of elapsed time $\Delta t$ measured for our fiducial model.  Note that to be consistent with a normal distribution as discussed above, we have multiplied the standard deviation measured for particles with $e_0 = 0$ by a factor of $\sqrt{2 / (4 - \pi)}$.  As in the case of radial drift discussed in \S\ref{SS:rd}, little difference exists between particles with different initial eccentricities and the results scale linearly with the dimensionless parameter $\xi$, which measures the strength of the disk gravity.  Chi-square fitting with a $\xi\Delta t^{1/2}$ dependence to the growth in eccentricity deviation in all six models leads to (Fig.~\ref{Fi:det})
\begin{equation} \label{E:de}
  \sigma(\Delta e) = 
  (4.1\pm0.6)\times10^{-4}\,\xi\left(\frac{H}{R}\right)
                            \left(\frac{\Delta t}{P}\right)^{1/2}.
\end{equation}
Note that at 1~AU in an MMSN disk, $\xi \simeq 1$, $H / R \simeq 0.1$, and $P = 1$~yr, and thus the increase in eccentricity deviation due to hydromagnetic turbulence only amounts to about 0.04 in 1~Myr.

As with the case of radial drift discussed in \S\ref{SS:rd}, our measured spread of eccentricity in our fiducial model is in approximate agreement with OIM07.  On the other hand, N05 reported a typical eccentricity growth of $e \simeq 0.03$ for $\Delta t \sim 100P$.  With $\xi \simeq 6.6$ at radius $R = 2.7R_1$ in the model studied by N05 (Richard~P. Nelson 2009, private communication) and $H / R \simeq 0.1$, the eccentricity deviation given by equation~\eqref{E:de} is about one order of magnitude less than what was reported by N05.

One possible explanation for this inconsistency could be the presence of large-scale structures, particularly $m=1$ modes, in global, but not in local models.  OIM07 reported that the inclusion of an $m = 1$ mode in their torque formula induces a ten times greater impact of hydromagnetic turbulence on particle orbits, using the calibration with global MHD simulations provided by LSA04.  Inspection of LSA04, as well as N05, shows that the density structures in their global models often extend more than $\pi / 2$ in azimuthal angle, leading to the $m = 1$ mode having the largest amplitude.  Such large-scale modes indeed can be excited in self-gravitating disks, and they could also be excited by local turbulence. This idea is offered some support by the modest growth in density perturbations in larger boxes that we find (Fig.~\ref{Fi:mrit_box}), and the growth in $\alpha$ reported in larger boxes by \citet{JYK09}.

\begin{figure}[!tbp]
\begin{center}
\plotone{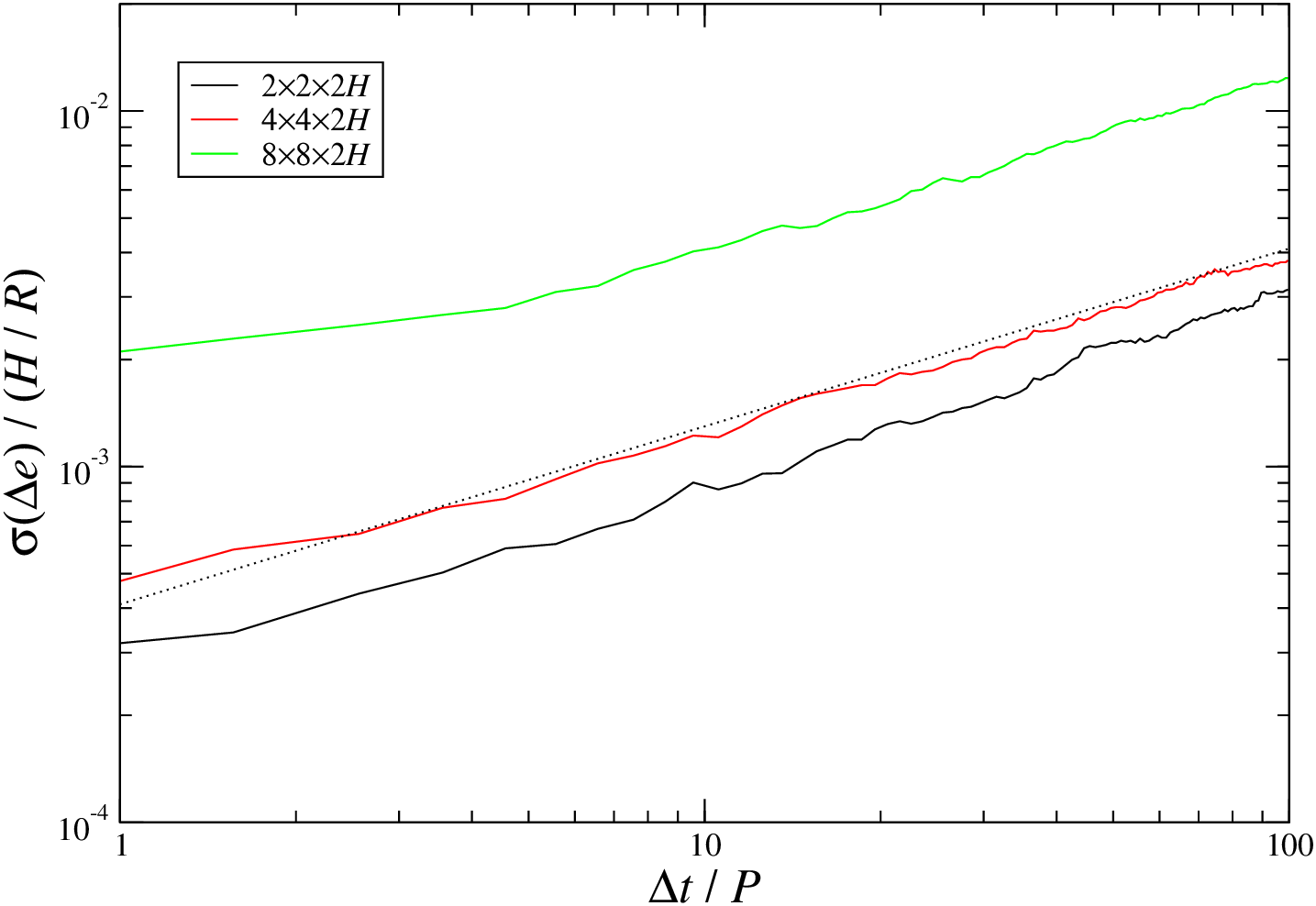}
\caption{Standard deviation of eccentricity deviation $\sigma(\Delta e)$ as a function of elapsed time $\Delta t$ for three different box sizes at a resolution of 32~points per scale height $H$ (\emph{solid} lines), where only the low-mass disk model ($\xi = 1$) and particles with zero initial eccentricity are considered.  For comparison, the straight \emph{dotted} line gives the best fit to the high-resolution model shown in Figure~\ref{Fi:det} (eq.~[\ref{E:de}]).}
\label{Fi:det_box}
\end{center}
\end{figure}

In Figure~\ref{Fi:det_box} we examine the growth in eccentricity deviation as a function of box size in our local models.  Increasing the horizontal box size by a factor of four indeed increases the eccentricity deviation by a factor of four, though this still does not account for the order of magnitude higher value found by N05 in his global model.  If $m = 1$ modes indeed can be similarly excited by well resolved turbulence, though, this could offer an explanation for the discrepancy.

However, using local models with mean azimuthal fields, \citet{GG09} reported that the coherent structures induced by the MRI are localized, with correlation lengths of about $0.05H$, $0.32H$, and $0.05H$ in radial, azimuthal, and vertical directions, respectively.  Therefore, it is possible that current global models might not have enough resolution to model such fine structures, which were then spuriously connected into extended structures resembling $m = 1$ modes.  To determine the physical reality of large-scale structure in turbulent disks, global models capable of resolving localized structures will be needed.

  \subsection{Horizontal Velocity Dispersion} \label{SS:hvd}

\begin{figure}[!htbp]
\begin{center}
\plotone{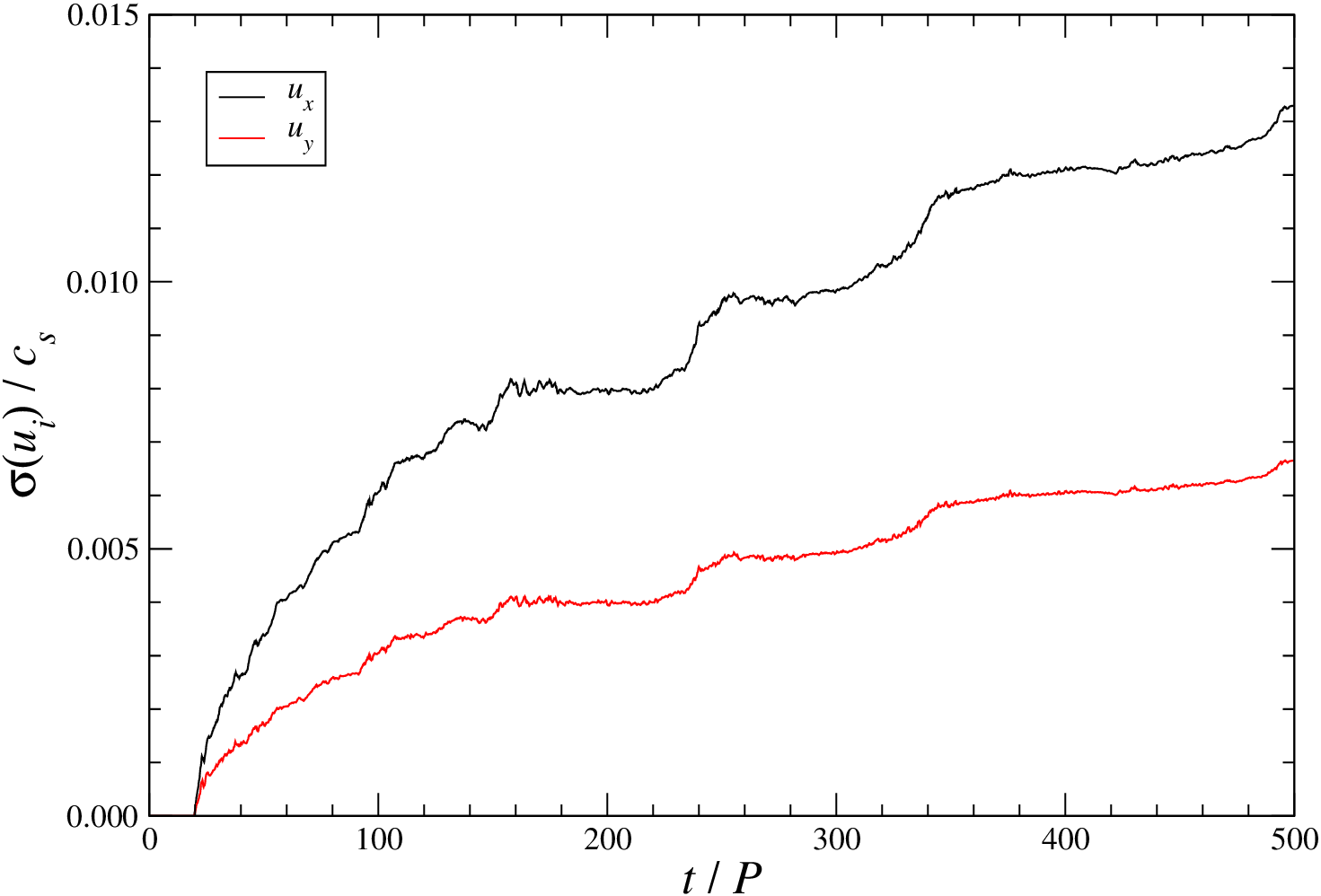}
\caption{Radial and azimuthal components of velocity dispersion $\sigma(u_x)$ and $\sigma(u_y)$ as a function of time $t$ for particles with initial eccentricity $e_0 = 0$ moving in the low-mass disk version of our fiducial model.  They are presented as running averages over one orbital period $P$.}
\label{Fi:uil0}
\end{center}
\end{figure}

Another important quantity in the study of planetesimal dynamics is the velocity dispersion of the particles.  We calculate the radial and the azimuthal components of velocity dispersion by taking the standard deviations of radial and azimuthal velocities for all particles, i.e., $\sigma(u_x)$ and $\sigma(u_y)$, respectively.  Since all the dynamical equations are linearized in terms of $x / R$ in the local shearing box approximation, the background shear flow is uniform irrespective of position, so the velocity dispersion for all particles in the computational domain is a well-defined local quantity.  Figure~\ref{Fi:uil0} plots $\sigma(u_x)$ and $\sigma(u_y)$ measured for particles with zero initial eccentricity in the low-mass disk version of our fiducial model as a function of time $t$.  The velocity dispersion monotonically increases with time, and thus hydromagnetic turbulence tends to steadily heat up a planetesimal disk.  Note that $\sigma(u_y) \sim \sigma(u_x) / 2$, a sanity check that our results are consistent with a swarm of non-interacting particles moving epicyclically in a Keplerian disk.

\begin{figure}[!htbp]
\begin{center}
\plotone{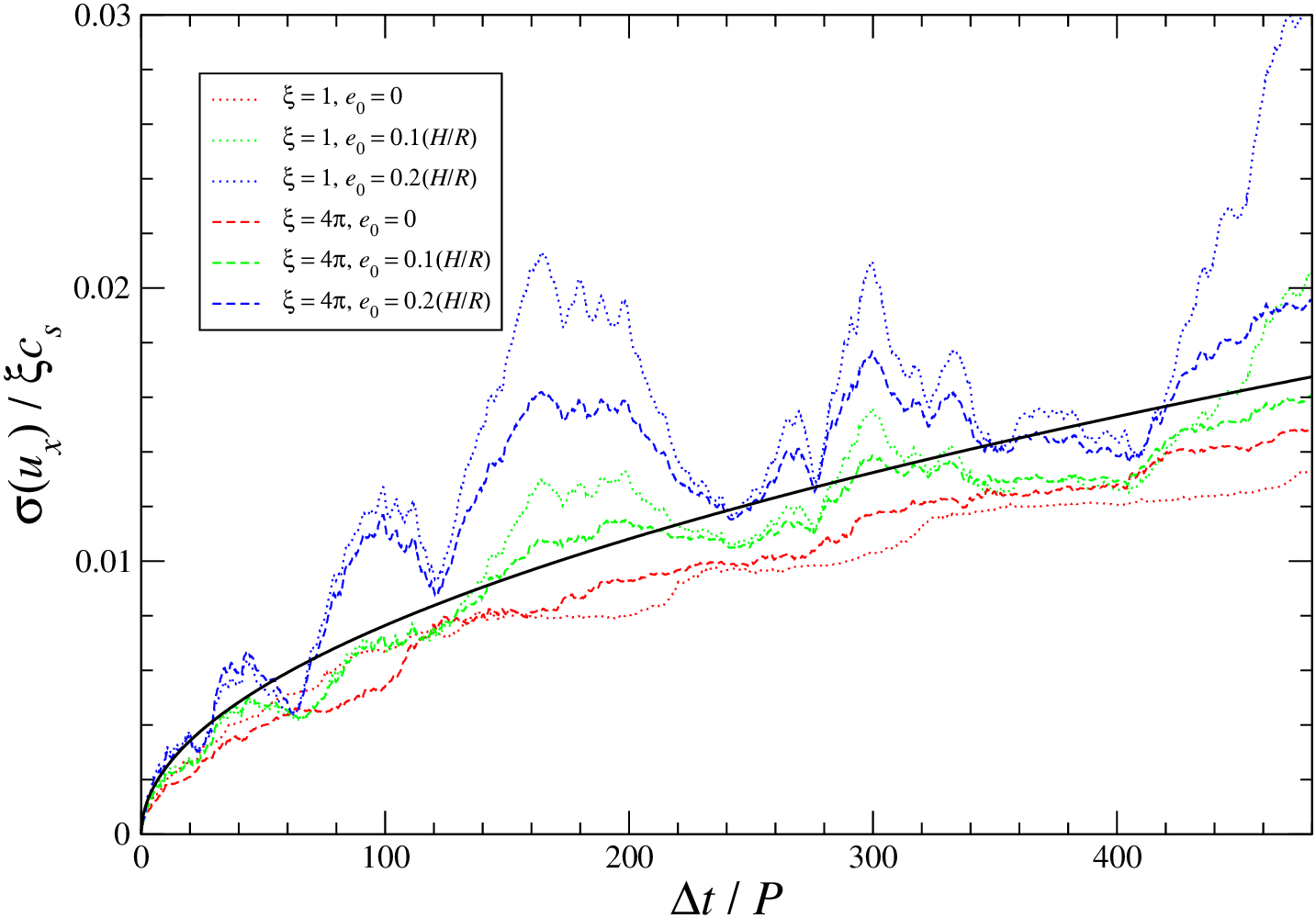}
\caption{Radial component of velocity dispersion $\sigma(u_x)$ as a function of elapsed time $\Delta t$ in our fiducial model, computed as running averages over one orbital period $P$.  The \emph{dotted} lines are obtained from low-mass disks ($\xi = 1$) while the \emph{dashed} lines are from high-mass disks ($\xi = 4\pi$).  Particles with initial eccentricities $e_0 = 0$, $0.1(H/R)$, and $0.2(H/R)$ are denoted by \emph{red}, \emph{green}, and \emph{blue} lines, respectively.  The \emph{solid black} line is the best fit to all six curves, assuming no explicit dependence on $e_0$.}
\label{Fi:ux}
\end{center}
\end{figure}

\begin{figure}[!htbp]
\begin{center}
\plotone{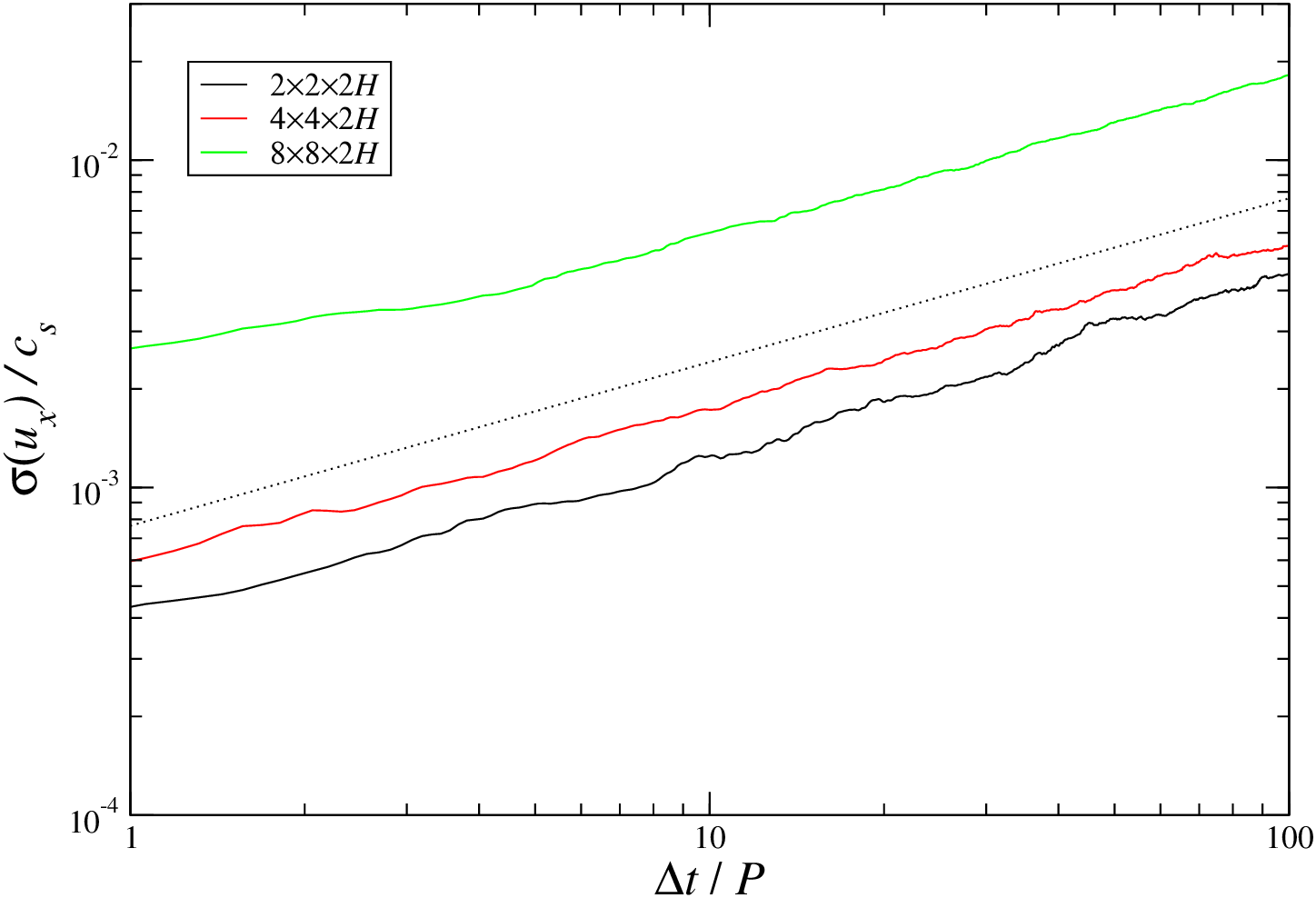}
\caption{Radial component of velocity dispersion $\sigma(u_x)$ as a function of elapsed time $\Delta t$ for three different box sizes at a resolution of 32~points per scale height $H$ (\emph{solid} lines), where only the low-mass disk model ($\xi = 1$) and particles with zero initial eccentricity are considered. For comparison, the straight \emph{dotted} line gives the best fit to the high-resolution model shown in Figure~\ref{Fi:ux} (eq.~[\ref{E:rvd}]).}
\label{Fi:ux_box}
\end{center}
\end{figure}

Figure~\ref{Fi:ux} compiles the radial velocity dispersions $\sigma(u_x)$ as a function of elapsed time $\Delta t$ measured in our fiducial model for particles with different initial eccentricity $e_0$ moving in disks of different disk gravity.  As is the case with the radial drift and eccentricity deviation discussed in \S\ref{SS:rd} and \S\ref{SS:ecc}, the results depend linearly on $\xi$, the strength of disk gravity, though there seems to be a slightly enhanced effect for particles with $e_0 > 0$.  Given the uncertainty involved in the numerical simulations, we assume it is a secondary effect as a first approximation.  The best fit to all six models is then given by
\begin{equation} \label{E:rvd}
  \sigma(u_x) = (7.6\pm1.6)\times10^{-4}\,\xi c_s
                \left(\frac{\Delta t}{P}\right)^{1/2}.
\end{equation}
The corresponding timescale $\tau_T$ for turbulent excitation of velocity dispersion can be estimated by $\tau \equiv \sigma / (\mathrm{d}\sigma / \mathrm{d}\Delta t)$, and we find from equation~\eqref{E:rvd}
\begin{equation} \label{E:t1}
  \tau_T = 
  \left(3.4\times10^6\,P\right)\xi^{-2}\left[\frac{\sigma(u_x)}{c_s}\right]^2.
\end{equation}
Increasing the box size increases the magnitude of this effect, as shown in Figure~\ref{Fi:ux_box}, possibly even non-linearly.

As noted in \S\ref{SS:ecc}, hydromagnetic turbulence can act either to excite the eccentricities of planetesimals or to circularize their eccentric orbits.  However, a perfectly cold disk of planetesimals with nonzero initial eccentricity but vanishing velocity dispersion will be monotonically heated by the turbulence, as shown above.  The velocity dispersion increases with time while the mean eccentricity may remain unchanged unless a significant fraction of particles are circularized.  Therefore, the eccentricity may not necessarily be proportional to the velocity dispersion for a swarm of planetesimals moving through hydromagnetic turbulence.

The velocity dispersion of a planetesimal disk also grows with time due to mutual gravitational scattering.  The corresponding timescale for a swarm of identical particles of mass $m_p$ can be estimated by \citep[e.g.,][]{PT06}
\begin{equation} \label{E:gs1}
  \tau_{GS} = \frac{\sigma^3(u_x)}{8\sqrt{\pi}G^2 m_p^2 n_p \ln\Lambda_p}
              \left[\frac{\sqrt{3}}{4}
                    \ln\left(\frac{2 + \sqrt{3}}
                                  {2 - \sqrt{3}}\right) - 1\right]^{-1},
\end{equation}
where $n_p$ is the number density of planetesimals and $\Lambda_p = 3\sigma^2(u_x) H_p / 4G m_p$, in which $H_p \sim \sqrt{2}\sigma(u_z) / \Omega_K$ is the scale height of the planetesimal disk determined by the vertical velocity dispersion $\sigma(u_z)$.  To find $\tau_{GS}$ in physical units, we assume for simplicity that most of the solid material in a protoplanetary disk is concentrated in planetesimals, and thus $n_p m_p \sim \varepsilon\rho_0 c_s / \sigma(u_z)$ where $\varepsilon$ is the solid-to-gas ratio.  We also assume $\sigma(u_z) \sim \sigma(u_x) / 2$, so that equation~\eqref{E:gs1} becomes
\begin{equation} \label{E:gs2}
  \tau_{GS} \simeq 
  \frac{\sqrt{\pi} P^2 \sigma^4(u_x)}
       {4\varepsilon \xi c_s G m_p \ln\Lambda_p}
  \left[\frac{\sqrt{3}}{4}
        \ln\left(\frac{2 + \sqrt{3}}{2 - \sqrt{3}}\right) - 1\right]^{-1}
\end{equation}
with
\begin{equation}
  \Lambda_p \simeq
  \frac{3\sqrt{2}P\sigma^3(u_x)}{16\pi G m_p}.
\end{equation}
We further focus our discussion on a velocity scale of order $v_\mathrm{esc}$, the escape velocity at the surface of a planetesimal:
\begin{equation}
  v_\mathrm{esc} = \left(\frac{32\pi}{3} G^3 m_p^2 \rho_p\right)^{1/6},
\end{equation}
where $\rho_p$ is the material density of the planetesimal.  This scale is of critical interest for planetary cores to accrete solid material; particles with relative velocities of order $v_\mathrm{esc}$ are more likely to coalesce into larger bodies than to be eroded into smaller pieces.  By assuming $\sigma(u_x) \sim v_\mathrm{esc}$, the timescales for heating a planetesimal disk by hydromagnetic turbulence and gravitational scattering become
\begin{equation}
  \tau_T = 
  \left(1.1\times10^2~\textrm{yr}\right)
  \xi^{-2}
  \left(\frac{\rho_p}{3~\textrm{g cm}^{-3}}\right)^{1/3}
  \left(\frac{m_p}{10^{18}~\textrm{g}}\right)^{2/3}
  \left(\frac{c_s}{10^5~\textrm{cm s}^{-1}}\right)^{-2}
  \left(\frac{P}{\textrm{yr}}\right)
\end{equation}
and
\begin{equation}
  \tau_{GS} \simeq
  \left(\frac{1.4\times10^5}{\ln\Lambda_p}~\textrm{yr}\right)
  \xi^{-1}
  \left(\frac{\varepsilon}{0.01}\right)^{-1}
  \left(\frac{\rho_p}{3~\textrm{g cm}^{-3}}\right)^{2/3}
  \left(\frac{m_p}{10^{18}~\textrm{g}}\right)^{1/3}
  \left(\frac{c_s}{10^5~\textrm{cm s}^{-1}}\right)^{-1}
  \left(\frac{P}{\textrm{yr}}\right)^2
\end{equation}
with
\begin{equation}
  \Lambda_p \simeq
  6.9\times10^3\,
  \left(\frac{\rho_p}{3~\textrm{g cm}^{-3}}\right)^{1/2}
  \left(\frac{P}{\textrm{yr}}\right),
\end{equation}
respectively.

With the scales assumed above and at 1~AU in an MMSN disk, $\tau_{GS} \sim 2\times10^4$~yr while measurement from our high-resolution fiducial model described by equation~\eqref{E:t1} gives $\tau_T \sim 100$~yr.  For larger objects approaching the planetary mass regime, with $m_p = 0.001M_\earth$, $\tau_{GS} \sim \tau_T \sim 3\times10^6$~yr.  Therefore, hydromagnetic turbulence probably dominates the heating of a disk of kilometer-sized planetesimals, while gravitational scattering may be more important for objects approaching Earth size.  Models with larger boxes yield smaller values of $\tau_T$, strengthening this conclusion.  Note that $\tau_{GS}$ increases more steeply with $\sigma(u_x)$ and $P$ but decreases less rapidly with $\xi$ than $\tau_T$ (eqs.~[\ref{E:t1}] and [\ref{E:gs2}]), and thus hydromagnetic turbulence  gains dominance over gravitational scattering for larger velocity dispersion, larger distance to the host star, and more massive gas disks than assumed here.

\section{IMPLICATIONS FOR PLANET FORMATION} \label{S:ipf}

In this section, we apply our results on the orbital evolution of zero-mass particles to two specific problems in planet formation.  First, we estimate the strength of diffusive migration of protoplanets due to hydromagnetic turbulence and discuss their survivability under type~I migration following the analytical framework established by JGM06.  Secondly, we revisit the proposition of IGM08 that planetesimals may suffer from collisional destruction as a result of hydromagnetic turbulence excitation of velocity dispersion among them.
  
  \subsection{Diffusive Migration of Protoplanets} \label{SS:dmp}

Using a Fokker-Planck formalism, JGM06 derived an advection-diffusion equation to describe the evolution of the distribution of protoplanets under the influence of both type~I migration and hydromagnetic turbulence.  \citet{AB09} further consolidated the analysis by studying more realistic disk density structure with both spatial and time dependence.  These authors found that turbulence tends to reduce the lifetimes of most protoplanets while allowing some of them to linger long enough to survive rapid inward type~I migration.  The likelihood of producing a planetary system with a specific configuration sensitively depends on the strength of the diffusive migration induced by the turbulence, however.  JGM06 and \citet{AB09} calibrated the turbulence strength with the global disk models computed by LSA04, \citet{NP04}, and N05.  In contrast, using a local disk model, OMM07 found that the strength might be several orders of magnitude less than what was estimated by JGM06.  In this section, we provide a new assessment based on the disk models studied in this work.

As shown in \S\ref{SS:rd}, an initial delta function in the distribution of mean orbital radii of zero-mass particles is spread with time into a normal distribution of constant mean.  A $t^{1/2}$ time dependence for the standard deviation of the distribution suggests that this process can be described by a diffusion equation of the form
\begin{equation} \label{E:diffus}
  \frac{\partial f}{\partial t} = \frac{\partial}{\partial x}
  \left(\mathcal{D}\frac{\partial  f}{\partial x}\right),
\end{equation}
where $\mathcal{D}$ is the diffusion coefficient and $f = f(t,x)$ is the distribution function: $f(t,x)\mathrm{d}x$ is the probability of finding a particle with a radial displacement in $(x, x + \mathrm{d}x)$ at time $t$.  Let us write,
\begin{equation} \label{E:normal}
  f(x,t) = \frac{1}{\sigma(t)\sqrt{2\pi}}
           \exp\left[-\frac{x^2}{\sigma^2(t)}\right],
\end{equation}
\begin{equation} \label{E:sigma1}
  \sigma(t) = \sigma_1\left(\frac{t}{P}\right)^{1/2} \qquad (t > 0),
\end{equation}
where $\sigma_1$ is a proportionality constant.  Substituting equation~\eqref{E:normal} into equation~\eqref{E:diffus}, we find that $\mathrm{d}\sigma / \mathrm{d}t = 2\mathcal{D} / \sigma$ and thus with $\sigma(t \rightarrow 0^+) = 0$,
\begin{equation} \label{E:sigma2}
  \sigma(t) = 2\mathcal{D}^{1/2} t^{1/2}.
\end{equation}
By comparing equations~\eqref{E:sigma1} and \eqref{E:sigma2}, the diffusion coefficient $\mathcal{D}$ is related to $\sigma_1$ and $P$ by
\begin{equation}
  \mathcal{D} = \frac{\sigma_1^2}{4P}.
\end{equation}
The best-fit to our fiducial high resolution, small box model, given by equation~\eqref{E:dx}, can be substituted here to find
\begin{equation} \label{E:dcr}
  \mathcal{D}(R) = 3.6\times10^{-8}\,\xi^2\left(\frac{H^2}{P}\right)
\end{equation}
by identifying $t$ with $\Delta t$.   In this derivation, we have assumed that the stochastic torque exerted by hydromagnetic turbulence is a local process such that the diffusion coefficient $\mathcal{D}$ is sufficiently constant near $x = 0$.

We are now in a position to estimate the diffusion coefficient $D(J)$ of JGM06 in comparison to $\mathcal{D}(R)$, where $J = m_p\left(G M_\star R\right)^{1/2}$ is the orbital angular momentum of a protoplanet of mass $m_p$ orbiting a star of mass $M_\star$ on a quasi-circular orbit at a radial distance $R$.  Since $D(J)$ and $\mathcal{D}(R)$ have dimensions of $[J^2 / t]$ and $[R^2 / t]$, respectively, they may be related by $D(J) \sim \mathcal{D}(R) (\partial J / \partial R)^2 = (J/2R)^2 \mathcal{D}(R)$.  Using dimensional arguments, JGM06 defined a dimensionless parameter $\epsilon$ to describe the uncertainties associated with hydromagnetic turbulence:
\begin{equation} \label{E:dcj}
  D(J) = (2.1 \times 10^{-3})
  \epsilon(2\pi)^3 \frac{\Sigma^2 J^7}{G^2 M_\star^4 m_p^5} =
  \frac{(2.1 \times 10^{-3})}{16\pi}\epsilon\xi^2
  \left(\frac{H}{R}\right)^2\left(\frac{J^2}{P}\right).
\end{equation}
By comparing equations~\eqref{E:dcr} and \eqref{E:dcj}, we find that
\begin{equation}
  \epsilon\simeq2.2\times10^{-4} \label{E:eps}
\end{equation} 
for our fiducial model.  Note that $\epsilon$ is a constant independent of $R$ given our scalings, in agreement with the assumption of constant $\epsilon$ made by JGM06.

\begin{figure}[!htbp]
\begin{center}
\plotone{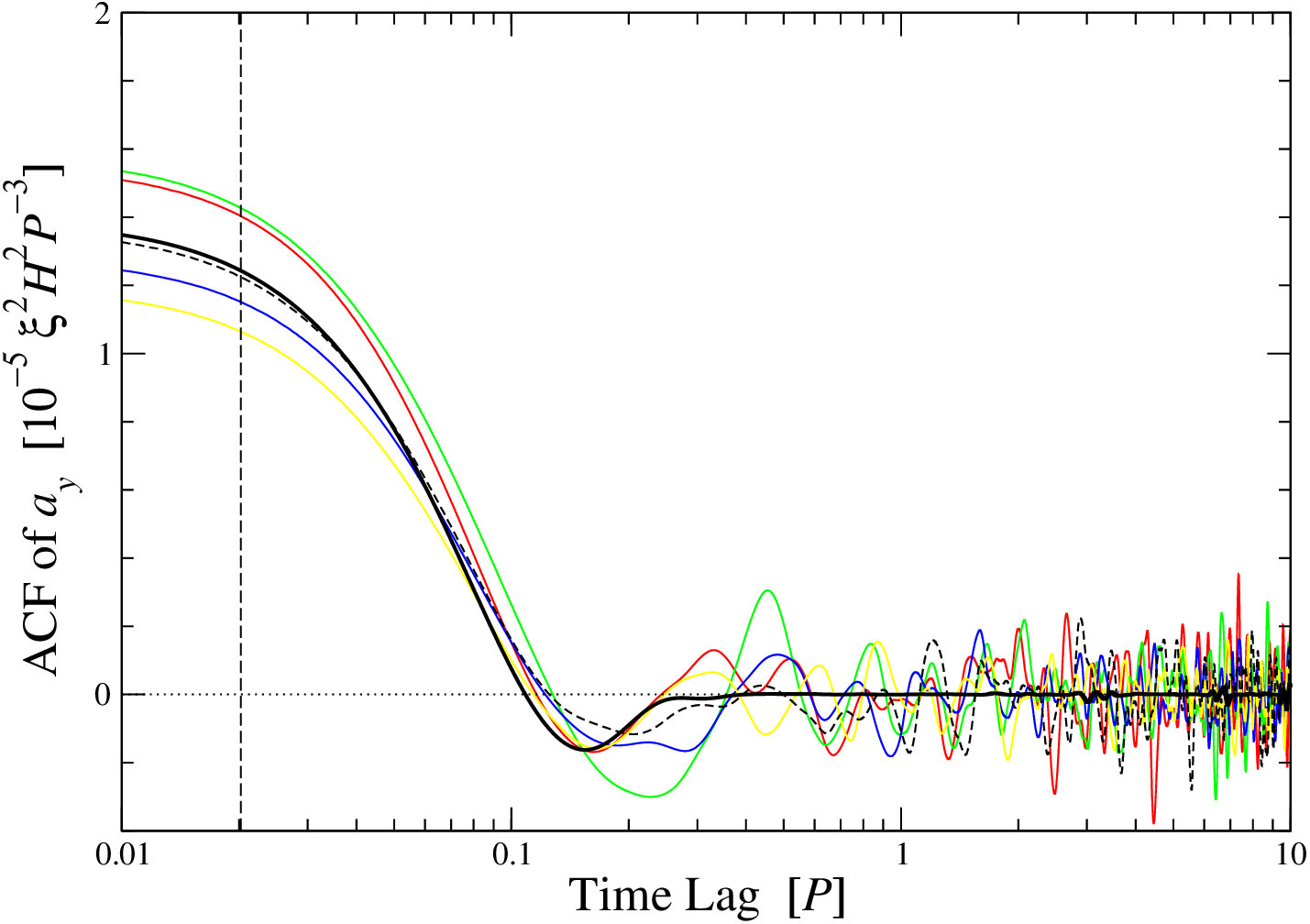}
\caption{Autocorrelation functions of the azimuthal (torque) component of the gravitational force per unit mass, $a_y$, exerted by the turbulent gas in our fiducial model.  The \emph{short-dashed black} curve corresponds to $a_y$ calculated at the center of the box, while the various \emph{colored} curves correspond to $a_y$ calculated following the orbits of several randomly selected particles.  The average over all particles is given by the \emph{solid black} line, which shows almost no power beyond the second zero-crossing.  The vertical \emph{long-dashed black} line indicates our estimated correlation time of the torques using equation~\eqref{E:ct}.  Notice the negative part of the ensemble-averaged autocorrelation function beyond the first zero-crossing, which may be responsible for reducing the diffusion coefficient (see \S\ref{SS:dmp}).}
\label{Fi:acf}
\end{center}
\end{figure}

The value of $\epsilon$ estimated in equation~\eqref{E:eps} is about an order of magnitude smaller than what was reported by OMM07 for a stratified disk model with zero net flux on a $1\times4\times4H$ grid at the same resolution as our fiducial model with 64 points per scale height.  To identify the reason for this discrepancy, we further evaluate the magnitude and the correlation time of the torques exerted by the turbulent gas on the particles in our model.  Firstly, the root-mean-square of the $y$-component of the gravitational force per unit mass exerted by the gas over all time and particles is $a_{y,\mathrm{rms}} \simeq (3.7\times10^{-3}) \xi H P^{-2} = (4.2\times10^{-3})(2\pi G\Sigma)$.  We find negligible difference between torques calculated following the particles along their orbits and those calculated at the fixed center of the box.  The magnitude we obtained is reasonably consistent with the value of $a_{y,\mathrm{rms}} = (3.2\times10^{-3})(2\pi G\Sigma)$ reported by OMM07 at the same numerical resolution.  Secondly, we plot in Figure~\ref{Fi:acf} the autocorrelation functions (ACF) of $a_y$ for several randomly selected particles as well as at the center of the box.  The results are again similar to what was reported by OMM07, indicating a similar estimate for the correlation time, $\tau_c$.  The source of the discrepancy appears to be neither of these factors.

We notice, though, that OMM07 could have overestimated the correlation time by equating it to the value of the second zero-crossing of the ACF.  Strictly speaking, the correlation time should instead be computed by integrating an ensemble average of the ACF over all possible realizations, as motivated by the definitions of the diffusion coefficient and the correlation time in JGM06:
\begin{equation} \label{E:jgmd}
  D(J) \equiv
  \frac{1}{2}\int_{-\infty}^{\infty}
  \overline{\delta\Gamma(t - \frac{\tau}{2}, J)
            \delta\Gamma(t + \frac{\tau}{2}, J)}\,
  \mathrm{d}\tau
\end{equation}
and
\begin{equation} \label{E:jgmt}
  \tau_c \equiv D(J) / \overline{\delta\Gamma^2(t,J)},
\end{equation}
where $\delta\Gamma(t,J)$ is the fluctuating part of the torque and the overline denotes ensemble average. Since we distribute numerous particles uniformly over the entire computational domain, the ACF obtained for each particle can be considered as one realization, and the average of the ACFs over all particles may resemble the true ensemble average.  This averaged ACF is shown by the solid black curve in Figure~\ref{Fi:acf}.  Note that the oscillation occurring at time lag longer than the second crossing for each particle is much reduced, indicating the noise nature of the autocorrelation at long time lag.  However, the negative value of the ACF between the first two zero-crossings remains significant.  This interval represents the anti-diffusion nature of the stochastic torques that we believe could be responsible for reducing the diffusion coefficient.  Therefore, we suggest that a better approximation for the correlation time in accordance with equations~\eqref{E:jgmd} and~\eqref{E:jgmt} be
\begin{equation} \label{E:ct}
  \tau_c \approx
  \frac{\int_0^{\infty} \overline{\mathrm{ACF}(\tau)}\,\mathrm{d}\tau}
       {2\overline{\mathrm{ACF}(0)}},
\end{equation}
where $\overline{\mathrm{ACF}(\tau)}$ is the ensemble-averaged ACF of the torque per unit mass $a_y$ as a function of time lag $\tau$.  Using equation~\eqref{E:ct}, we find that in our fiducial model $\tau_c \simeq 0.020P$, about one order of magnitude less than the value of $\tau_c \simeq 0.31P$ reported by OMM07.  With the same approximation $D(J) \simeq m_p^2 R^2\overline{a_y^2}\tau_c$ used by OMM07, we find that $\epsilon \simeq 1.7\times10^{-4}$, in good agreement with the estimate of $\epsilon\simeq2.2\times10^{-4}$ we derived from our direct measurement of particle radial drifts.

Therefore, we have achieved consistent results using two independent approaches to estimating the diffusion coefficient, one by direct measurement as in equation~\eqref{E:dcr} and the other by the definition of correlation time given by JGM06 (eq.~[\ref{E:jgmt}]).  In principle, our direct measurement should be robust while using $D(J) \simeq \tau_c \overline{\delta\Gamma^2}$ can only be considered as an approximation.  To use the latter approach, one should refer to equation~\eqref{E:ct} to measure the correlation time of the stochastic torques, instead of conventional methods like using the zero-crossing of the ACF or the timescale of the peak of the temporal power spectrum, which probably give about one order of magnitude larger values.  It will be enlightening for studies based on global models to conduct the same exercise described in this section, since this will likely explain part of the orders of magnitude discrepancy in diffusion coefficients derived from local versus global models.  (Although the box size effects we have identified must also contribute to this discrepancy, they appear unable to increase the derived value of $\epsilon$ by more than one order of magnitude.)

According to Figures~6 and~7 of JGM06, the value of $\epsilon$ we inferred from our simulations indicates that advective (type~I) migration dominates over diffusive (stochastic) migration for the parameter space JGM06 have investigated.  For an Earth-mass protoplanet at a radial distance up to 100~AU in the MMSN disk and in a viscous disk with $\alpha = 0.02$, advection dominates when $\epsilon\la10^{-2}$ and $10^{-1}$, respectively.  For a protoplanet of mass as low as 0.01~$M_\earth$ at $R = 10$~AU in the same disk models, advection dominates when $\epsilon\la10^{-3}$ and $10^{-2}$, respectively.  The critical distance and mass for the transition between dominance of advection and diffusion given our small estimate of $\epsilon$ for our fiducial model is outside of the parameter space explored by JGM06, and even our largest box has a value of $\epsilon$ less than an order of magnitude larger.  By inspection of Figure~7 in JGM06, though, the transition masses for an object at 10~AU in the MMSN disk and in the viscous disk probably lies at about $10^{-3}$ and $10^{-4}~M_\earth$, respectively.  Therefore, our results suggest that hydromagnetic turbulence does not significantly affect the secular migration of Earth-sized protoplanets in regimes of current astrophysical interest.  Nevertheless, torques exerted by turbulent density perturbations seem to be a dominant agent determining the orbital dynamics of kilometer-sized planetesimals.
  
  \subsection{Collisional Destruction of Planetesimals} \label{SS:cdp}

IGM08 suggested that hydromagnetic turbulence may inhibit the growth of kilometer-sized planetesimals.  They argued that the velocity dispersion excited by the turbulence could be so large that collisions between planetesimals exceed their material strength or self-gravity, leading to destruction.  Their conclusion, however, relies on orbital integrations incorporating the heuristic, stochastic formulas for the time history of gravitational torques provided by LSA04, which in turn were calibrated using MHD simulations of a global disk model.  As pointed out in \S\ref{SS:ecc}, a possible inconsistency exists between global and local models.  Since the latter shows a significantly lower effect on orbital dynamics of planetesimals, it is worthwhile revisiting the planetesimal growth problem in light of the results highlighted in this work.

For comparison purposes, we adopt the same scalable MMSN disk model as used by IGM08.  The gas density and the speed of sound in the mid-plane are given by
\begin{equation}
  \rho_0 =
  \left(1.8\times10^{-9}~\textrm{g cm}^{-3}\right)
  f_g\left(\frac{R}{\textrm{AU}}\right)^{-11/4}
\end{equation}
and
\begin{equation}
  c_s =
  \left(1.1\times10^5~\textrm{cm s}^{-1}\right)
  \left(\frac{R}{\textrm{AU}}\right)^{-1/4},
\end{equation}
respectively, where a solar-type host star ($M_\star = M_\sun$) is assumed and $f_g$ is a scale factor.  When $f_g = 1$, the disk mass is about 1.4 times that of an MMSN disk.  The corresponding $\xi$-parameter and ratio of disk scale height to radial distance then become
\begin{equation} \label{E:xi_mmsn}
  \xi = 1.5 f_g \left(\frac{R}{\textrm{AU}}\right)^{1/4}
\end{equation}
and
\begin{equation} \label{E:hor_mmsn}
  \frac{H}{R} =
  0.051\left(\frac{R}{\textrm{AU}}\right)^{1/4},
\end{equation}
respectively.  Substituting equations~\eqref{E:xi_mmsn} and \eqref{E:hor_mmsn} into equation~\eqref{E:de}, which is derived from our fiducial high-resolution model, we arrive at
\begin{equation} \label{E:de_mmsn}
  \sigma(\Delta e) =
  3.1\times10^{-5}\,f_g
  \left(\frac{R}{\textrm{AU}}\right)^{-1/4}
  \left(\frac{\Delta t}{\textrm{yr}}\right)^{1/2}.
\end{equation}
Comparing equation~\eqref{E:de_mmsn} with equation~(13) of IGM08 with the understanding that $\sigma(e) = \sigma(\Delta e)\sqrt{(4 - \pi) / 2}$ for particles with zero initial eccentricity (see \S\ref{SS:ecc}), we find the value of the dimensionless parameter $\gamma$ --- a measure of the strength of hydromagnetic turbulence used by IGM08 --- in our orbital integrations to be $\gamma \simeq 2.0\times10^{-4}$.  The rest of the analysis performed by IGM08 remains unchanged since the effects induced by hydromagnetic turbulence are all incorporated in the parameter $\gamma$.

As noted in \S\ref{SS:ecc}, enlarging the horizontal size of our shearing box by a factor of four at our medium resolution increases the amplitude of the eccentricity deviation by about a factor of four.  Nevertheless, the largest box we have studied gives $\gamma \simeq 6\times10^{-4}$, which remains somewhat smaller than the $\gamma \sim 10^{-3}$--$10^{-2}$ estimated by IGM08.

The rather small values of $\gamma$ we obtained indicate that hydromagnetic turbulence might not pose as serious a threat to the growth of kilometer-sized planetesimals as suggested by IGM08.  These authors compared the critical radii of planetesimals for accretive and erosive regimes due to different turbulence strengths $\gamma = 10^{-2}, 10^{-3}$, and $10^{-4}$ in their Figures~3, 4, and~5, respectively.  Therefore, the values of $\gamma$ we measured point to a scenario in between what is predicted by Figures~4 and~5 of IGM08.  In this scenario, the erosive regime only appears in the outer regions of a young protoplanetary disk and it disappears rapidly with decreasing disk mass.  Kilometer-sized planetesimals may be able to evade collisional destruction in the inner regions of the disk.

We reiterate that the value of $\gamma$ measured here pertains to a local region with turbulent stresses such that $\alpha\sim10^{-2}$ and $\Delta\rho / \rho_0 \sim 10\%$.  As discussed in \S\ref{SS:ecc}, the validity of our results in a global context remains to be demonstrated by numerical experiments on a global disk model with a resolution capable of resolving the characteristic scales of coherent turbulence structures.

\section{CONCLUSIONS} \label{S:conc}

In this work, we have used local, shearing-box simulations to study the dynamics of massless planetesimals in a turbulent, isothermal, unstratified, gas disk driven by the MRI.  With a uniform, vertical magnetic field but without explicit physical dissipation, the saturated turbulence is maintained at a roughly constant level, showing convergence with increasing resolution.  By adopting a suitable magnitude for the net magnetic flux, we produce a fiducial disk model with turbulent accretion at the level of $\alpha \sim 10^{-2}$ and with root-mean-square density perturbations $\Delta\rho / \rho_0 \sim 10$\%.  As discussed in \S\ref{SS:cusr}, this model can be scaled to other physical systems of interest, as long as the assumption of negligible self-gravity of the gas remains valid.  After the hydromagnetic turbulence in our fiducial model reaches saturation, we distribute numerous particles of zero mass and integrate their orbital motion under the gravitational influence of the turbulent gas.

The stochastic nature of the orbital evolution of these particles is evident, so we characterize their orbital dynamics with statistical distributions, finding three major results.  First, although the mean orbital radius does not change, particles slowly drift away from their original radii, so that the distribution of radii grows with time.  Second, gravitational force from density perturbations produced in hydromagnetic turbulence can both excite and damp the eccentricities of particle orbits, with again no change in the mean value, but a growing width of distribution if the particles possess non-negligible initial eccentricities.  Finally, the planetesimal disk is heated up by the turbulence, a process dominating over gravitational scattering between particles in most physical conditions relevant to protoplanetary gas disks and planetesimal sizes.  A corollary of these results is that eccentricity does not serve as a good indicator of the velocity dispersion of the particles.

The amplitude of orbital changes driven by the turbulence in our local models is significantly smaller than what was reported in recent global models (LSA04; N05; IGM08).  Two possible explanations for this discrepancy suggest themselves: either insufficient resolution in the global models or the lack of convergence with box size in our local shearing box model, as discussed in \S\ref{SS:ecc}.  If our local results are valid, they indicate that although hydromagnetic turbulence can drive radial diffusion, eccentricity variations, and relative velocities of planetesimals and protoplanets, these effects may not be dominant in determining their evolution.  In particular, it appears that type~I migration dominates over turbulent radial drift for objects well above $10^{-4}$~M$_\earth$.  In addition, hydromagnetic turbulence might not be exciting sufficient velocity dispersion to drive planetesimals into an erosive regime that would inhibit their further growth.  Before these results can be considered robust, however, it will be necessary to elucidate, and hopefully reconcile, the differences that have appeared between global and local models.

\acknowledgments
We thank Jeffrey~S.\ Oishi and Richard~P.\ Nelson for the clarification of their works and useful discussions.  We thank the anonymous referee for urging us to explore models with larger box sizes.  The computations reported here were performed on the Columbia and Pleiades systems of the NASA Advanced Supercomputing (NAS) Division.  This research is supported by the NASA Origins of Solar Systems Program under grant NNX07AI74G.

\appendix
\section{VELOCITY OF A PARTICLE AT THE APOGEE OF ITS ORBIT} \label{S:vpao}

In this section, we (re-)derive the velocity of a particle at the apogee of its elliptical orbit in the local shearing box approximation of a Keplerian disk.  We repeat the equation of motion~\eqref{E:eomp2} for a single particle without the gravity of the gas here:
\begin{eqnarray}
  \frac{\mathrm{d}u_x}{\mathrm{d}t} &=& 2\Omega_K u_y,\label{E:eompax}\\
  \frac{\mathrm{d}u_y}{\mathrm{d}t} &=& -\frac{1}{2}\Omega_K u_x,
    \label{E:eompay}
\end{eqnarray}
where we have dropped the subscript $p$ for clarity.  Eliminating $u_y$ in equations~\eqref{E:eompax} and \eqref{E:eompay} leads to
\begin{equation}
  \frac{\mathrm{d}^2 u_x}{\mathrm{d}t^2} + \Omega_K^2 u_x = 0.
\end{equation}
By assuming the particle is at the apogee at $t = 0$, the solution for $u_x$ is\begin{equation} \label{E:ux}
  u_x = -A\sin \Omega_K t,
\end{equation}
where $A$ is the amplitude of the radial velocity.  Since $\mathrm{d}x / \mathrm{d}t = u_x$ (eq.~[\ref{E:eomp1}]), the radial oscillation is then
\begin{equation}
  \Delta x \equiv x - x_0 = \frac{A}{\Omega_K}\cos \Omega_K t,
\end{equation}
where $x_0$ is the radial position of the center of the orbit.  From equation~\eqref{E:eompay}, the corresponding azimuthal velocity relative to the background shear flow is
\begin{equation}
  u_y = -\frac{1}{2}A\cos \Omega_K t = -\frac{1}{2}\Omega_K\Delta x.
\end{equation}
Since the eccentricity $e$ of the orbit is related with the amplitude of the radial oscillation by $e \approx \Delta x(t = 0) / R$, where $R$ is the distance to the central object,
\begin{equation}
  u_y(t = 0) = -\frac{1}{2}R\Omega_K e
             = -\frac{1}{2}H\Omega_K\left(\frac{e}{H / R}\right),
\end{equation}
where $H$ is the disk scale height.  This is just the initial condition~\eqref{E:icp} we set out to prove.  Note that we have normalized $e$ by the ratio $H / R$.


\end{document}